\documentclass[a4p,11pt]{article}
\usepackage[margin=2cm]{geometry}

\usepackage[numbers]{natbib}

\usepackage{subfigure}
\usepackage{amsmath,amssymb,bm,graphicx}
\usepackage{color}
\usepackage{booktabs}
\usepackage[colorlinks=true]{hyperref}
\usepackage{algorithm}
\usepackage{textcomp}
\usepackage{multirow}
\usepackage{microtype}
\usepackage{algpseudocode}
\usepackage{amsthm}
\usepackage{placeins}

\newcommand{\tabref}[1]{Table~\ref{#1}} 
\newcommand{\figref}[1]{Fig.~\ref{#1}} 
\newcommand{\equref}[1]{Eq.~(\ref{#1})}
\newcommand{\secref}[1]{Section~\ref{#1}}
\DeclareMathOperator*{\minimize}{minimize}
\DeclareMathOperator*{\argmin}{argmin}
\DeclareMathOperator*{\maximize}{maximize}
\DeclareMathOperator*{\argmax}{argmax}

\begin{document}
%
%
%
\title{
Sparse stability diagrams of LSCF method via strategic pole destabilization using orthogonal matching pursuit\thanks{This manuscript is a preprint and has been submitted to a journal for possible publication.}
}  

\author{
Shogo Shimada\thanks{Department of Mechanical Engineering, 
Meiji University, Kawasaki, Kanagawa 214-8571, Japan.}
\and
Akira Saito\thanks{Corresponding author. Department of Mechanical Engineering, 
Meiji University, Kawasaki, Kanagawa 214-8571, Japan. 
E-mail: asaito@meiji.ac.jp} 
}

\date{}
\maketitle

\begin{abstract}
In various engineering fields including mechanical, aerospace, and civil engineering, the identification of modal parameters, including natural frequencies, damping ratios, and mode shapes, is crucial for determining the vibration characteristics of engineered structures. A common method for identifying the modal parameters of structures involves experimental modal analysis using frequency response functions (FRFs) obtained from forced vibration tests. The least squares complex frequency (LSCF) domain method is a widely-used frequency-domain curve-fitting method for the FRFs using the polynomials of high order, which can extract modal parameters with high accuracy. However, increasing the polynomial order tends to result in the generation of non-physical spurious poles that need to be eliminated from the stability diagrams. 
To overcome this issue, we propose a method that strategically destabilize the stable yet spurious poles of the characteristic polynomials by making their coefficients as sparse as possible, via orthogonal matching pursuit (OMP). 
This results in sparse stability diagrams because unstable poles can be eliminated from the diagrams. In this paper, the proposed method is first applied to a numerically-obtained FRFs of a rectangular plate using finite element model, and its validity is discussed. Then, the method is applied to experimentally-obtained FRFs of rectangular plates with low-damping and with high-damping. Furthermore, to confirm its applicability to industrial applications with realistic complexity, it has also been applied to the FRFs of the electric machine's stator core used for electric vehicles. Based on the results, we have confirmed that the spurious roots can be eliminated from the stability diagrams without compromising accuracy for the cases considered. 
\end{abstract}



%

\section{Introduction} \label{sec:introduction}
The identification of modal parameters including natural frequencies, damping ratios, and mode shapes is important in determining the vibration characteristics of engineered structures. A common method for identifying the modal parameters of a structure is the experimental modal analysis (EMA), which utilizes the frequency response functions (FRFs) obtained from forced vibration experiments. 

The methods of EMA can be roughly divided into time-domain methods and frequency-domain methods. 
Representative time-domain EMAs include least-squares complex exponential method and Ibrahim time domain method~\cite{Ewins2009, SaitoKuno2020}. Time-domain methods are suitable for special cases with very low natural frequencies~\cite{Ewins2009}. 

On the other hand, frequency domain methods are also commonly used for the EMA, and one of the most widely-used frequency-domain modal parameter extraction methods are based on the least squares complex frequency domain (LSCF) method \cite{AuweraerEtAl2001,PeetersEtAl2004}. 
In this approach, the measured FRF is approximated using a fraction of polynomials. 
The polynomial order for approximation is set by the users a priori, which needs to be much larger than the expected number of modes in the frequency range of interest. 
Then, by solving the least squares (LS) problem, the coefficients of the polynomial approximation are determined such that the difference between the measured FRF and the polynomial approximation is minimized. 
Successively applying the LS for the increasing polynomial orders, stable poles that exist in the frequency range are determined. By solving the resulting characteristic equation derived from the obtained coefficients, natural frequencies and the corresponding damping ratios are extracted. 
However, in many cases, increasing the polynomial order results in the generation of non-physical poles that appear in the stability diagrams, which need to be manually removed by inspection of the users. 
Additionally, the presence of measurement noises can result in producing poles that may or may not physically exist~\cite{VerbovenEtAl2005}. Therefore, when performing modal analysis on complex structures that result in many natural frequencies in wide frequency range, the results often depend on the user and may require engineers with advanced expertise. To overcome these challenges, research has been conducted to automate the process of experimental modal analysis using various methods, such as clustering~\cite{TavaresEtAl2023}
, meta-heuristic approach~\cite{SitarzEtAl2016,SitarzEtAl2019} and Bayesian optimization~\cite{EllingerEtAl2023}.
In this study, we propose a method to overcome these challenges. Specifically, we propose a new algorithm that strategically destabilize the stable yet spurious non-physical poles in the stability diagrams by incorporating the orthogonal matching pursuit (OMP)~\cite{MallatZhang1993,PatiEtAl1993} into the LSCF method. This results in the sparse coefficients of the polynomial approximation. This approach achieves a reduction in the model complexity. To demonstrate the effectiveness of the proposed method, we have applied the method to both numerical and experimental data and the results are discussed in detail. 
The structure of this paper is stated as follows. In \secref{sec:theory}, we provide an overview of the conventional LSCF method and introduce the idea of OMP to augment the conventional LSCF method. In \secref{sec:numerical}, the proposed method is applied to numerical data and its validity is examined. In \secref{sec:study1}-\secref{sec:study3}, three case studies with experimental data are shown. 
In \secref{sec:study1}, the case with small damping is examined using a flat plate made of aluminum alloy. 
In \secref{sec:study2}, the case with large damping is examined using a flat plate made of natural rubber. 
In \secref{sec:study3}, the case with industrial application is shown using a motor stator for the traction drive of electric vehicles. 
Lastly, in \secref{sec:con}, we conclude our findings.
\section{Mathematical formulation} \label{sec:theory}
\subsection{Least squares complex frequency domain method}
First, let us restrict our attention to the cases where FRFs are obtained for a single input and $n_o$ outputs, and denoted as 
\begin{equation}
\hat{\bf H}(\omega)=[\hat{H}_1(\omega),\dots,\hat{H}_{n_p}(\omega)]^{\rm T}, 
\end{equation} 
and $\hat{H}_o(\omega)\in\mathbb{C}$ for $o=1,\dots,n_p$. 
In the LSCF method with a common denominator model, the FRF is approximated by using fractions of polynomials as follows.
\begin{equation}
{\bf H}({\omega})={\bf B}({\omega})/A({\omega}),\label{eq1}
\end{equation}
where ${\bf H}({\omega}), {\bf B}({\omega})\in\mathbb{C}^{n_o}$, $A({\omega})\in\mathbb{C}$, ${N_i}$ is the number of inputs, and $n_o$ is the number of measurement DOFs, and ${\bf H}(\omega)=[H_1(\omega),\dots,H_{n_p}(\omega)]^{\rm T}$. 
With the LSCF method, ${\bf B}({\omega})$ and ${A}({\omega})$ are assumed to be expressed as polynomials. Namely, denoting the $o$th component in ${\bf B}(\omega)$ as $B_o(\omega)$, 
\begin{equation}
H_o(\omega)=B_o(\omega)/A(\omega)\label{eq11}
\end{equation}
where $B_o({\omega}) ={\bf b}_o^{\rm{T}}\bm{\Omega}({\omega})$, $A({\omega})={\bf a}^{\rm{T}}\bm{\Omega}({\omega})$, and $\bm{\Omega}(\omega)$ is a vector of basis functions whose components are written as $\bm{\Omega}({\omega})=[{\Omega}^{0},{\Omega}^{{1}}({\omega}),{\Omega}^{{2}}({\omega})\dots,{\Omega}^{{n_{p}}}({\omega})]^{\rm T}$ where ${\Omega}({\omega})={\rm e}^{{-{\rm j}}{\omega}{T_s}}$, ${T_s}$ is the sampling period and ${n_{p}}$ is the maximum polynomial order that is set by the user. ${\bf b}_o$ and ${\bf a}$ are the vectors of coefficients for the basis functions, i.e., 
${\bf b}_o= [{b_{o,0}},{b_{o,1}}, \dots,{b_{o,n_{p}}} ]^{\rm T}$, ${\bf a}  = [{a_0},{a_1},\dots,{a_{n_{p}}}   ]^{\rm T}$. 
Note that ${\bf b}_o$ for $o=1,\dots,n_o$ and ${\bf a}$ are concatenated and treated as the vector of unknown coefficients as 
$\bm{\theta}=[{\bf b}^{\rm T}_1,\dots,{\bf b}^{\rm T}_{n_o},{\bf a}]^{\rm T}$. 

Next, we consider that measured FRFs are fit by the mathematical model described above. Assume that the FRFs are measured at $n_f$ discrete frequency lines $\omega_k$, $k=1,\dots,n_f$. 
If the unknown coefficients $\bm{\theta}$ are determined such that the errors between $\hat{H}_o(\omega_k)$ and $H_o(\omega_k,\bm{\theta})$ for all $o=1,\dots,n_o$ and $k=1,\dots,n_f$ are minimized, it is expected that the resulting ${\bf H}(\omega,\bm{\theta})$ is a good approximation of $\hat{\bf H}(\omega)$. 
Such a minimization problem is defined as follows. 
Defining a weighted and linearized error as 
\begin{align}
\epsilon_o({\omega_k,\bm{\theta}})
&= w({\omega_k})
\left[
\left\{{\bf b}_o^{\rm T}\bm{\Omega}(\omega_k)
\right\}
-
\left\{
{\bf a}^{\rm T}\bm{\Omega}(\omega_k)
\right\}{\hat{H}_o}({\omega_k})
\right],
\label{eq:error}
\end{align}
where $w(\omega_k)$ is a frequency-dependent weighting function, which is typically computed from measured coherence function associated with $\hat{H}_o(\omega)$. 
Now defining $\bm{\epsilon}(\bm{\theta})\triangleq[\bm{\epsilon}_1^{\rm T},\dots,\bm{\epsilon}_{n_o}^{\rm T}]^{\rm T}$, where $\bm{\epsilon}_o=[\epsilon_o(\omega_1,\bm{\theta}),\dots,\epsilon_o(\omega_{Nf},\bm{\theta})]^{\rm T}$, we find $\bm{\theta}$ that minimizes the squared norm of the weighted errors, i.e., 
\begin{align}
\minimize_{\bm{\theta}}\mathcal{L}(\bm{\theta}):=
\frac{1}{2}
\lVert \bm{\epsilon}(\bm{\theta})\rVert^2_2.\label{eq2}
\end{align}
The minimizer of Eq.~\eqref{eq2} satisfies $\frac{\partial\mathcal{L}}{\partial\bm{\theta}}={\bf 0}$, i.e., ${\bf J}\bm{\theta}={\bf 0}$, which results in the corresponding normal equation, 
\begin{equation}
\left({\bf J}^{\rm H}{\bf J}\right)\bm{\theta}={\bf 0}, \label{eq3}
\end{equation}
where ${\bf J}$ is the Jacobian of $\mathcal{L}(\bm{\theta})$ with respect to $\bm{\theta}$, or ${\bf J}=\frac{\partial\mathcal{L}}{\partial\bm{\theta}}$, and $^{\rm H}$ denotes the conjugate transpose of a matrix. Equation~\eqref{eq3} can be re-written as an expanded form as
\begin{equation}
 \begin{bmatrix}
  {\bf R}_1 & {\bf 0} & \cdots & {\bf 0}& {\bf S}_1\\
  {\bf 0} & {\bf R}_2 & \cdots & {\bf 0}& {\bf S}_2\\
  \vdots& \vdots& \ddots& \vdots& \vdots\\
  {\bf 0} & {\bf 0}& \cdots & {\bf R}_{n_o} & {\bf S}_{n_o}\\
  {\bf S}_1^{\rm H}& {\bf S}_2^{\rm H}& \cdots& {\bf S}^{\rm H}_{n_o}&
  \sum^{n_o}_{o=1}{\bf T}_o
 \end{bmatrix}
 \begin{bmatrix}
  {\bf b}_1\\ {\bf b}_2\\ \vdots \\{\bf b}_{n_o}\\ {\bf a}
 \end{bmatrix}
={\bf 0}\label{eq4}
\end{equation}
where ${\bf R}_o$, ${\bf S}_o$ and ${\bf T}_o$ for $o=1,\dots,n_o$ are submatrices whose $r$-$s$ components are written as:
\begin{align}
 \left({\bf R}_o\right)_{rs}&
 =\sum_{f=1}^{n_f}w^2(\omega_f)\left\{\Omega(\omega_f)^{r-1}\right\}^{\rm
 H}\Omega(\omega_f)^{s-1},\\
 \left({\bf S}_o\right)_{rs}&
 =-\sum_{f=1}^{n_f}w^2(\omega_f)\hat{H}_o(\omega_f)\left\{\Omega(\omega_f)^{r-1}\right\}^{\rm H}\Omega(\omega_f)^{s-1},\\
 \left({\bf T}_o\right)_{rs}&
 =\sum_{f=1}^{n_f}w^2(\omega_f)\left|\hat{H}_o(\omega_f)\right|^2\left\{\Omega(\omega_f)^{r-1}\right\}^{\rm H}\Omega(\omega_f)^{s-1}.
\end{align}
By expressing ${\bf b}_o$ for $o=1,\dots{n_o}$ in Eq.~\eqref{eq4} using ${\bf a}$, 
\begin{equation}
 {\bf b}_o=-{\bf R}_o{\bf S}_o{\bf a},\quad o=1,\dots{n_o}.\label{eq5}
\end{equation}
Substituting Eq.~\eqref{eq5} back into Eq.~\eqref{eq4}, we obtain 
\begin{equation}
{\bf C}
{\bf a}={\bf 0}
\end{equation}
where 
\begin{equation}
{\bf C}\triangleq\left\{
\sum^{n_o}_{o=1} \left({\bf T}_o-{\bf S}_o^{\rm H}{\bf R}_o^{-1}{\bf S}_o
\right)\right\}
\end{equation}
and ${\bf C}\in\mathbb{C}^{(n_p+1) \times (n_p+1)}$, ${\bf a}\in\mathbb{C}^{(n_p+1) \times 1}$. 
It is known that when the coefficient of the highest-order term of ${\bf a}$ is set to 1, the physical poles tend to become stable, while mathematical spurious poles become unstable. As a result, the distinction between physical and mathematical poles becomes easier~\cite{CaubergheEtAl2005}. Namely, if $a_{n_p}$ is set to 1, 
\begin{equation}
\begin{bmatrix}
C_{0,0}&\dots&C_{0,n_p-1}&C_{0,n_p}\\
            &&\vdots&\\
C_{n_p,0}&\dots&C_{n_p,n_p-1}&C_{n_p,n_p}            
\end{bmatrix}
\begin{bmatrix}
a_0\\
\vdots\\
a_{n_p-1}\\
1
\end{bmatrix}
={\bf 0}, 
\end{equation}
Discarding the last equation $C_{n_p,0}a_0+\dots+C_{n_p,n_p}=0$ and re-arranging the terms yield the system of equations with $n_p$ equations for $n_p$ unknowns, 
\begin{equation}
\begin{bmatrix}
C_{0,0}&\dots&C_{0,n_p-1}\\
            &\vdots&\\
C_{n_p-1,0}&\dots&C_{n_p-1,n_p-1}
\end{bmatrix}
\begin{bmatrix}
a_0\\
\vdots\\
a_{n_p-1}\\
\end{bmatrix}
= -
\begin{bmatrix}
C_{0,n_p}\\
\vdots\\
C_{n_p-1,n_p}
\end{bmatrix}. 
\label{eq6}
\end{equation}
where $C_{i,j}$ denotes the $i$-$j$ component of ${\bf C}$. 
Or in a compact notation, 
\begin{equation}
{\bf D}{\bf x}={\bf d}. \label{eq7}
\end{equation}
where ${\bf x}=[a_0,\dots,a_{n_p-1}]^{\rm T}$, ${\bf D}$ is the upper left submatrix of {\bf C}, ${\bf d}$ is the rightmost column of 
${\bf C}$ negated, and ${\bf D}\in\mathbb{C}^{n_p\times n_p}$, ${\bf d}\in\mathbb{C}^{n_p \times 1}$. 
Note that both ${\bf D}$ and ${\bf d}$ consist of known quantities from measurements and the polynomial model expressed in Eq.~\eqref{eq11}. 
Solving Eq.~\eqref{eq7} with respect to ${\bf x}$ gives ${\bf a}$ of $A(\omega)$ because ${\bf a}=[{\bf x}^{\rm T},1]^{\rm T}$. Since ${\bf a}$ is a vector of the polynomial coefficients of $A(\omega)$, the roots of the polynomial can be obtained by solving the eigenvalue problem of the companion matrix of ${\bf a}$. 
Denoting the obtained $r$th eigenvalue of $A(\omega)$ as $z_r$, since $z_r={\rm e}^{-\lambda_r T_s}$, $\lambda_r=-\log(z_r)/T_s$. Therefore, considering that 
$\lambda_r=-\zeta_r\omega_r\pm{\rm j}\omega_r\sqrt{1-\zeta_r^2}$, 
$f_{d_r}=\omega_r\sqrt{1-\zeta_r^2}/2\pi={\rm Im}\left(\lambda_r\right)/2\pi$ where $f_{d_r}$ is the $r$ th damped natural frequency. Moreover, 
$\zeta_r=-{\rm Re}\left(\lambda_r\right)/|\lambda_r|$ is the $r$th modal damping ratio, and $f_r=|\lambda_r|/2\pi$ is the $r$th undamped natural frequency. 

Typically, $n_p$ is set much larger than the number of physical modes that exist in the frequency range of interest because we do not know how many modes exist in the range a priori. Therefore, the eigenvalue problem of the companion matrix corresponding to Eq.~\eqref{eq7} typically produce many mathematical poles that do not correspond to any natural frequencies that physically exist. To distinguish such mathematical poles from the real physical poles, Eq.~\eqref{eq7} is solved repeatedly with the increasing polynomial order, and a {\it stability diagram} is drawn to see the stability of the obtained poles, which may guarantee the existence of the pole in the real system. 
Namely, denoting the lower right $i\times i$ submatrix of ${\bf D}$ at the $i$th iteration as ${\bf D}_{i}$, bottom sub-vectors of ${\bf x}$ and ${\bf d}$ of size $i$ as ${\bf x}_{i}$ and ${\bf d}_{i}$, respectively, the following linear system of equations are solved to obtain part of ${\bf a}$, i.e., 
\begin{equation}
{\bf D}_{i}{\bf x}_{i}={\bf d}_{i},\quad{i=1,\dots,n_p-1}. \label{eq8}
\end{equation}
where ${\bf x}_1=[a_{n_p}]$ and ${\bf x}_i=[a_{n_{p}-i+1},\dots,a_{n_{p}}]^{\rm T}$ for $i=2,\dots,n_p$. 
Once ${\bf x}_i$ is obtained, the roots of the corresponding characteristic polynomial $A(\omega)$ and the eigenvalues $z_r$ are obtained, which follows the calculations of $\zeta_r$ and $f_r$. This process is repeated until $i=n_p$. 
The overall process of the conventional LSCF method is summarized in Algorithm~\ref{alg1}. 
{\color{black}{It is noted that the maximum polynomial order ($n_p$) should be chosen such that it does not fall below the number of modes in the frequency range, which is unknown. Therefore, it is recommended to set the maximum polynomial order as large as possible, yet it should be small enough not to produce too many spurious poles.}}
\begin{algorithm}[!tbp]
\normalsize{
\caption{Conventional Least Squares Complex Frequency Domain method}\label{alg1}
\begin{algorithmic}[1]
\State Compute ${\bf C}$ from measured FRFs and coherence
\State Extract ${\bf D}$ and ${\bf d}$ from ${\bf C}$
\State Set $n_p$
\For {$i=1$ to $n_p$}
\State ${\bf D}_i$ $\gets$ lower right $i\times i$ submatrix of ${\bf D}$ 
\State Compute ${\bf x}_i$ such that ${\bf D}_i{\bf x}_i={\bf d}_i$
\State Compute roots $z_r$ of $A(\omega)$ using ${\bf x}_i$
\State $\lambda_r\gets-\ln(z_r)/T_s$
\State $f_r\gets|\lambda_r|/2\pi$
\State $\zeta_r\gets-{\rm Re}(\lambda_r)/|\lambda_r|$
\State Draw $f_r$'s on the stability diagram for the polynomial order of $i$
\EndFor
\State Extract stable $f_r$ and $\zeta_r$ from the stabilization diagram
\State Compute mode shapes using $f_r$ and $\zeta_r$
\end{algorithmic}
}
\end{algorithm}

\subsection{Sparse modeling by destabilization of stable spurious poles} 
Even though the constraint on $a_{n_p}=1$ is effective in distinguishing mathematical and physical poles, solving Eq.~\eqref{eq8} and the resulting characteristic polynomials still produces poles that may or may not physically exist especially when dealing with noisy measurement data. Even with the use of the stability diagram, determining the validity of such poles is still difficult. Hence, we try to reduce the number of such poles while maintaining the accuracy and detectability of the major modes of interest, which results in a clearer stability diagram where the important modes can be easily found. 

It is known that the roots of polynomials can be very sensitive to small perturbations in their coefficients~\cite{Wilkinson1959}. 
{\color{black}{In general, the LSCF method uses polynomials with larger number of poles than that of the actual poles. In other words, the coefficient vector ${\bf x}_i$ contains redundant coefficients. This naturally yield mathematical spurious poles that are the byproduct of the overfitting of the noisy data due to the redundant coefficients. Therefore, it is expected that the spurious poles can be perturbed by removing such redundant coefficients in the polynomial. 
On the other hand, the poles that actually exist in the system are dominated by the resonant characteristics manifested in the FRFs and are not greatly affected by the measurement noises in comparison with the spurious poles. Thus, it is expected that perturbing, or even removing the coefficients in ${\bf x}_i$ does not greatly affect the dominant poles. The spurious poles, however, are dominated by the measurement noises and by perturbing the redundant coefficients, they may be perturbed to an extent that they cross the stability boundary. 
}}
By exploiting such characteristics, the proposed method attempts to intentionally {\it destabilize} the stable yet spurious poles while keeping the dominant poles of the characteristic polynomial, 
i.e.,the number of coefficients are reduced as much as possible. 

Namely, the number of stable spurious poles in the stability diagram are reduced as follows. 
The stable pole is defined as the one with positive damping ratio, or $\zeta_r>0$. In other words, ${\rm Re}(\lambda_r)<0$ because $\zeta_r=-{\rm Re}\left(\lambda_r\right)/|\lambda_r|$. The unstable pole, on the other hand, is defined as the one with negative damping ratio, or $\zeta_r<0$. In other words, ${\rm Re}(\lambda_r)>0$. Now considering the following relationship
\begin{equation}
{\rm Re}(\lambda_r)=
{\rm Re}\left(-\log(z_r)/T_s\right)=-\ln(|z_r|)/T_s, 
\end{equation}
we can see that $\ln(|z_r|)<0\Leftrightarrow|z_r|<1$. This means that ${\rm Re}(\lambda_r)>0$ and $\lambda_r$ is an unstable pole. Or, if $\ln(|z_r|)>0\Leftrightarrow|z_r|>1$, which means that ${\rm Re}(\lambda_r)<0$ and $\lambda_r$ is a stable pole. 
This implies that if we can increase the number of poles with $|z_r|<1$, the number of unstable poles increases. 
To this end, we hypothesize that by enforcing the distribution of the terms in the characteristic polynomial to be sparse while keeping the highest order of the polynomial, the resulting poles of the polynomial tend to become unstable, which result in sparse stability diagram. 

\begin{figure}[!tbp]
\centering
\subfigure[100 out of 100 nonzero coefficients ($69\%$ of the roots are inside the unit circle)]{\includegraphics[scale=.8]{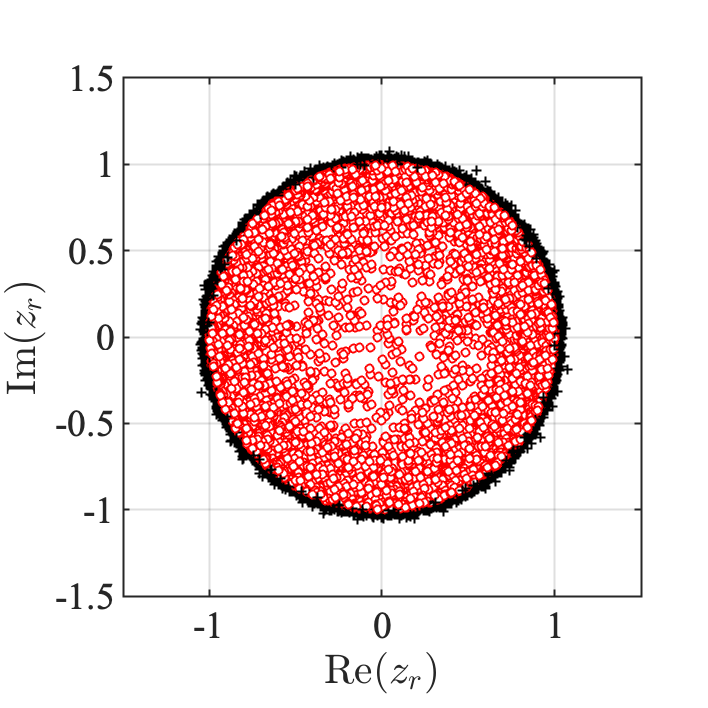}}
\hspace{2em}
\subfigure[70 out of 100 nonzero coefficients ($70\%$ of the roots are inside the unit circle)]{\includegraphics[scale=.8]{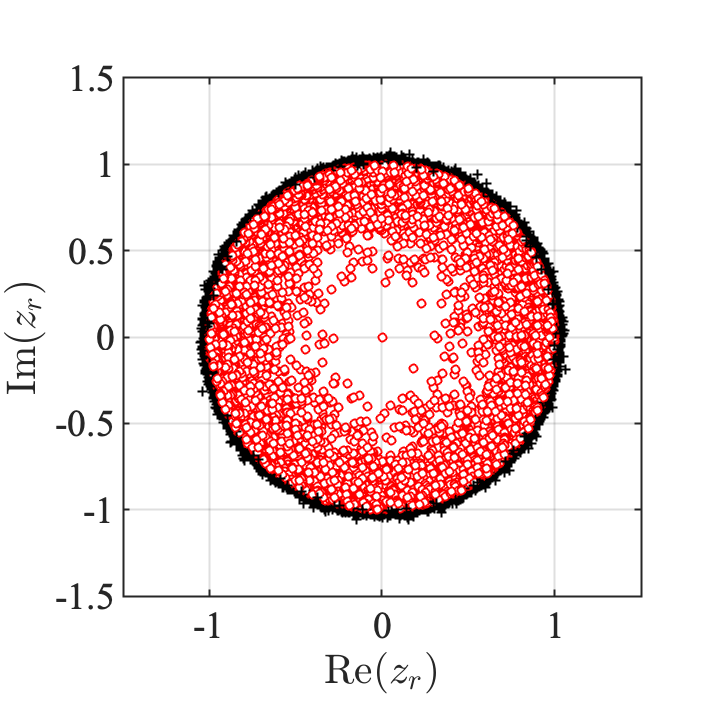}}
\\
\subfigure[30 out of 100 nonzero coefficients ($76\%$ of the roots are inside the unit circle)]{\includegraphics[scale=.8]{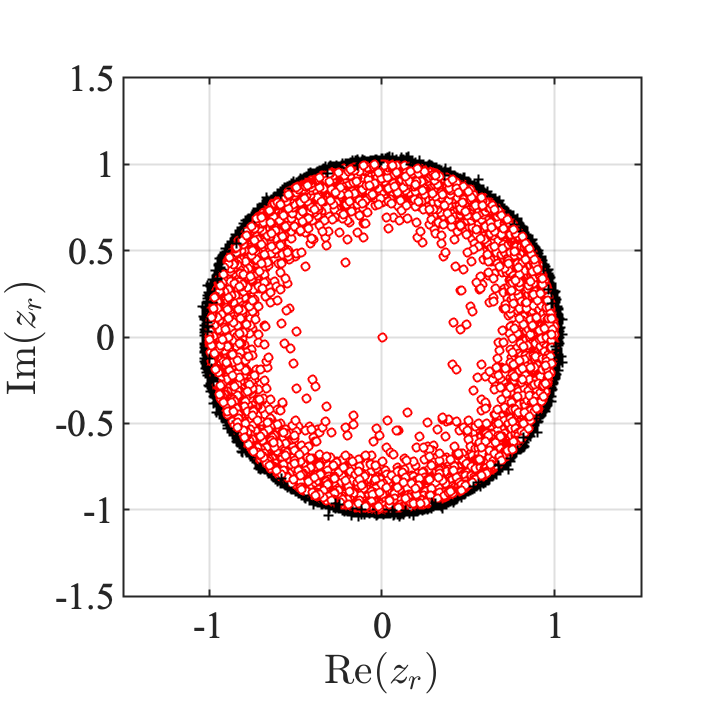}}
\hspace{2em}
\subfigure[5 out of 100 nonzero coefficients ($96\%$ of the roots are inside the unit circle)]{\includegraphics[scale=.8]{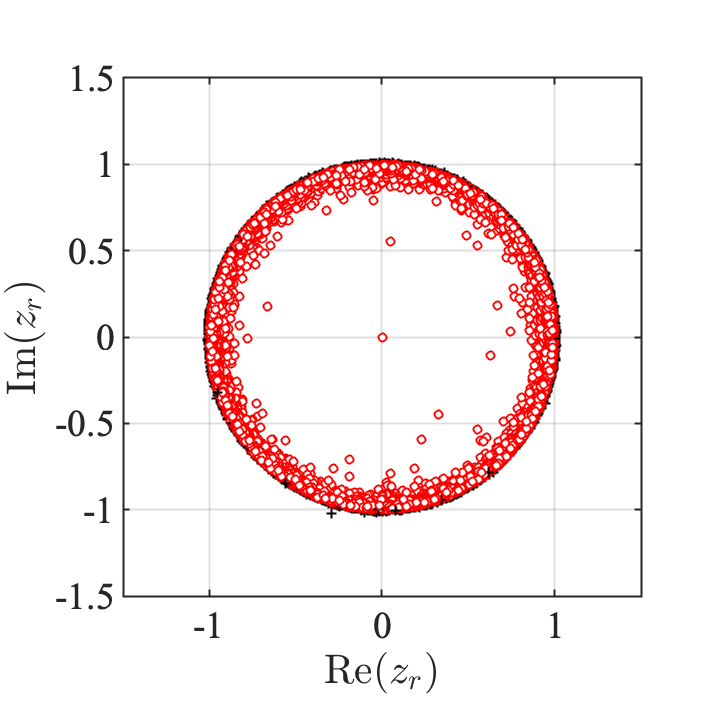}}
\caption{Roots of polynomials of order 100 for various numbers of nonzero coefficients randomly chosen for 1,000 trials. ${\color{black}{\circ}}$: inside the unit circle (unstable), $+$: outside the unit circle (stable)}\label{sparsity:circle}
\end{figure}

\begin{figure}[!tbp]
\centering
\includegraphics[scale=1]{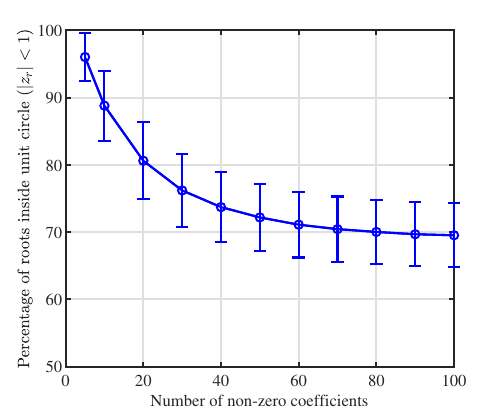}
\caption{Relationship between the percentage of the roots of the polynomial with $|z_r|<1$ and the number of coefficients kept in the polynomial while keeping the highest degree term. The errorbar indicates the standard deviation.}
\label{sparsity:trend}
\end{figure}

To illustrate this concept, a simple numerical example is considered. Let us now assume that we have a polynomial $f(z)$ of degree $n$. The coefficient of the highest order term is set to 1. Namely, 
\begin{equation}
f(z)=z^n + a_{n-1}z^{n-1}+\dots+a_0z^0, \label{eq:poly}
\end{equation}
where $a_{j}\in\mathbb{C},j=1,\dots,n-1$. Now consider a problem of finding the roots $z_r$ of Eq.~\eqref{eq:poly}, or $z=z_r$ that satisfies $f(z)=0$. 
To examine the effects of sparsifying $a_{j}$ on the stability of the roots, the roots of $f(z)=0$ were sought for randomly chosen $a_{j}$'s with pre-determined sparsity level. The degree of the polynomial was fixed to $n=100$, and the roots were obtained for the $a_{j}$'s whose real and imaginary parts were chosen from a uniformly distributed random numbers. 
The computation was repeated by keeping certain numbers of the dominant coefficients, where the degree of dominance was determined based on $|a_j|$. The set of computations were repeated for 1,000 times, and the results are shown in \figref{sparsity:circle}, where all of the obtained roots are plotted ($10^5$ roots per plot) for 100, 70, 30, and 5 nonzero coefficients. As can be seen, as the number of non-zero coefficients increases, the roots tend to gather near the unit circle in the complex plane. More importantly, as the number of non-zero coefficients increases, the number of roots with $|z_r|<1$ also increases. 
To see this trend more clearly, the relationship between the percentage of the roots of the polynomial with $|z_r|<1$ and the number of coefficients kept in the polynomial while keeping the highest degree term has been examined and plotted in \figref{sparsity:trend}. As can be seen, as the number of non-zero coefficients decreases, i.e., the sparsity level of the polynomial increases, the percentage of the roots with $|z_r|<1$ increases. 

If Eq.~\eqref{eq:poly} is regarded as the characteristic polynomial $A(\omega)$ in the LSCF formulation, the coefficients $a_j$'s correspond to ${\bf x}_i$ in Eq.~\eqref{eq8}. Therefore, from the observation above, we can deduce that if we are able to sparsify ${\bf x}_i$ when solving Eq.~\eqref{eq8}, the resulting characteristic polynomial has many roots that lie inside a unit circle in the complex plane. This results in increasing the number of unstable poles with negative damping ratios, which can be eliminated in the stability diagrams. 

The question we have to consider now is how we sparsify the coefficients ${\bf x}_i$ of the characteristic polynomial in an {\color{black}{optimal}} manner. 
In the proposed method, we apply the OMP when solving Eq.~\eqref{eq8} to enforce ${\bf x}_i$ as sparse as possible. 
\FloatBarrier
\subsubsection{Orthogonal matching pursuit} 
The OMP is an algorithm used in signal processing and statistics to find the best sparse approximation of a signal~\cite{WangEtAl2023}. 
The basic algorithm of the OMP is reviewed as follows. Consider the following regression problem: 
\begin{equation}
\bm{\Phi}{\bf x}={\bf y}\
\label{eq:regression_problem 2}
\end{equation}
where $\bm{\Phi}$ is a matrix consisting of known quantities, or a dictionary, ${\bf x}$ contains the unknowns to be determined, ${\bf y}$ is the vector of known data, and $\bm{\Phi}\in\mathbb{C}^{m \times n}$, 
 ${\bf x}\in\mathbb{C}^{n \times 1}$, ${\bf y}\in\mathbb{C}^{m \times 1}$, and $n\geqslant m$. 
Suppose that we are interested in finding ${\bf x}$ with a small number of non-zero elements, or a sparse solution of ${\bf x}$. 
This is achieved as follows. 

The first step is to find ${\bf x}$ with only a single non-zero element, which is referred to as 1-sparse. Such a solution of ${\bf x}$ at this step is designated as ${\bf x}_{[1]}$ and found as the minimizer of the following minimization problem: 
\begin{equation}
\minimize_{{\bf x}:\lVert{\bf x}\rVert_0=1}: \lVert
\bm{\Phi}{\bf x}-{\bf y}
\rVert^2_2\label{eq:min_prob}. 
\end{equation}
Namely, if the $i$th column of $\bm{\Phi}$ is denoted as $\bm{\phi}_i$, the index $i$ that achieves Eq.~\eqref{eq:min_prob} can be obtained by finding the column in $\bm{\Phi}$ that most resembles ${\bf y}$ in the sense of inner product, i.e., the index that satisfies the following maximization:
\begin{equation}
\maximize_{i}:\frac{|\langle\bm{\phi}_i,{\bf y}\rangle|}{\lVert\bm{\phi}_i\rVert_2}, \label{eq:min_prob2}
\end{equation}
where $\langle\cdot,\cdot\rangle$ denotes the inner product between two vectors. 
The index that achieves Eq.~\eqref{eq:min_prob2} is referred to as $i_{[1]}$ and the corresponding normalized inner product value is referred to as $x_{[1]}$, i.e., 
\begin{equation}
{x_{[1]}}=\frac{\langle{\bm{\phi}_{i_{[1]}}},{\bf y}\rangle}{||{\bm{\phi}_{i_{[1]}}}||_2^2}. 
\label{eq:soukan_x1}
\end{equation}
Denoting $x_{[1]}$ as the found non-zero element in ${\bf x}_{[1]}$, ${\bf x}_{[1]}=x_{[1]}{\bf e}_{i_{[1]}}$ where 
${\bf e}_i\in\mathbb{C}^{n\times 1}$ is a unit vector with 1 at the $i$th element and 0's at all the other elements. 
The residual between the data ${\bf y}$ and its 1-sparse representation of $\bm{\Phi}{\bf x}_{[1]}$ is then computed as 
\begin{equation}
{\bf r}_{[1]}={\bf y}-\bm{\Phi}{\bf x}_{[1]}
\label{eq:OMP_1}
\end{equation}
Note that the residual ${\bf r}_{[1]}$ is orthogonal to $\bm{\phi}_{i_{[1]}}$. 
Namely, ${\bf r}_{[1]}={\bf y}-\bm{\phi}_{i_{[1]}}x_{[1]}$. Therefore, 
$\langle{\bf r}_{[1]},\bm{\phi}_{i_{[1]}}\rangle=\langle{\bf y},\bm{\phi}_{i_{[1]}}\rangle-x_{[1]}\lVert \bm{\phi}_{i_{[1]}}\rVert_2^2=0$. 

The second step seeks for another column of $\bm{\Phi}$ that best approximates the residual ${{\bf r}_{[1]}}$. 
This can be obtained by finding the index $i$ that achieves
\begin{equation}
\maximize_{i}:\frac{|\langle\bm{\phi}_i,{\bf r}_{[1]}\rangle|}{\lVert\bm{\phi}_i\rVert_2}.
\label{eq:min_prob3}
\end{equation}
The index that achieves Eq.~\eqref{eq:min_prob3} is referred to as $i_{[2]}$. 
Then, defining $\bm{\Phi}'\triangleq[\bm{\phi}_{i_{[1]}},\bm{\phi}_{i_{[2]}}]$, the following least squares solution is sought: 
\begin{equation}
\minimize_{\bf x}:\lVert
\bm{\Phi}'{\bf x}-{\bf y}
\rVert^2_2. 
\end{equation}
The solution of this is denoted as $[x_{[1]},x_{[2]}]$, and 
the 2-sparse solution of ${\bf x}$, or ${\bf x}_{[2]}$ is defined as ${\bf x}_{[2]}={x_{[1]}}{{\bf e}_{i_{[1]}}}+{x_{[2]}}{{\bf e}_{i_{[2]}}}$. 
Note that $x_{[1]}$ has been updated and different from that obtained from the previous step. 
The residual is then updated as
\begin{equation}
{\bf r}_{[2]}={\bf y}-\bm{\Phi}{\bf x}_{[2]}. 
\end{equation}
This process is repeated for ${k}$ times. As a result, 
${k}$ indices are selected, i.e., $k$-sparse approximation of ${\bf y}$ with the dictionary $\bm{\Phi}$ can be obtained as 
\begin{equation}
{\bf x}_{[k]}={x_{[1]}}{{\bf e}_{i_{[1]}}}+{x_{[2]}}{{\bf e}_{i_{[2]}}}+\dots+{x_{[k]}}{{\bf e}_{i_{[k]}}}
\label{OMP_k}
\end{equation}

In the algorithm of the LSCF method, Eq.~\eqref{eq8} corresponds to Eq.~\eqref{eq:regression_problem 2}. With the OMP, we want to find a sparse ${\bf x}_i$ that approximately satisfies Eq.~\eqref{eq8}, which will become the coefficients of the characteristic polynomial with the expectation that this results in producing unstable poles while preserving the stability of the dominant poles.

\subsubsection{Determination of the sparsity} 

The number of OMP iterations, $k$, is determined by utilizing the relationship between the stability of ${L0}$ optimization and ${L1}$ optimization~\cite{EladEtAl2010}.
The purpose of the regression problem \equref{eq:regression_problem 2}  is to find ${\bf x}$ with a small number of non-zero elements. The number of these nonzero elements is expressed as the $L0$-norm of ${\bf x}$, or 
${\lVert{\bf x}\rVert_0}=\sum_{i=1}^{n}|x_i |^0$ and can be obtained as follows.
\begin{equation}
n_0=\minimize_{\bf x}{||{\bf x}||_0},\quad\mbox{subject to:}\quad
\bm{\Phi}{\bf x}={\bf y}\label{n0}
\end{equation}
Solving this $L0$ optimization problem is challenging because it requires searching through all possible combinations of zero and nonzero elements of ${\bf x}$, causing the exponential growth of the computational complexity. Therefore, an alternative approach is to replace the $L0$ norm with the $L1$ norm. Namely, ${\lVert{\bf x}\rVert_1=\sum_{i=1}^{n}{|x_i|^1}}$ and obtained as 
\begin{equation}
{\bf x}'=\argmin_{\bf x}{||{\bf x}||_1},\quad\mbox{subject to:}\quad
\bm{\Phi}{\bf x}={\bf y}\label{n1}
\end{equation}
and 
\begin{equation}
n_1 = \rVert{\bf x}'\rVert_0. 
\end{equation}
This is solvable with a reasonable amount of computational effort with existing methods. In the proposed algorithm, Least Absolute Shrinkage and Selection Operator~(LASSO)~\cite{Tibshirani1996,Tibshirani2011} is utilized. Namely, the following minimization problem is solved: 
\begin{equation}
\minimize_{\bf x}\quad\frac{1}{2}\lVert\bm{\Phi}{\bf x}-{\bf y}\rVert^2_2+\lambda\lVert{\bf x}\rVert_1\label{lasso}
\end{equation}
By solving Eq.~\eqref{lasso}, we expect to obtain a solution ${\bf x}$ that is not necessarily the sparsest but still sufficiently sparse. 
Since $n_0$ is the number of non-zero elements of the sparsest solution,  
the following inequality holds
\begin{equation}
n_0\leqslant n_1
\end{equation}
Therefore, the number of OMP iteration is set to $k=n_1$ to obtain sufficiently sparse solution. 
Based on the conventional LSCF algorithm shown in Algorithm~\ref{alg1}, the modified algorithm with the OMP is summarized in Algorithm~\ref{alg:cap}. 

\begin{algorithm}[!tbp]
\normalsize{
\caption{Least Squares Complex Frequency Domain method with OMP}\label{alg:cap}
\begin{algorithmic}[1]
\State Compute ${\bf C}$ from measured FRFs and coherence
\State Extract ${\bf D}$ and ${\bf d}$ from ${\bf C}$
\State $k\gets \min_{\bf x}\left\{\frac{1}{2}\lVert{\bf D}{\bf x}-{\bf d}\rVert^2_2+\lambda\lVert{\bf x}\rVert_1\right\}$
\State Set $n_p$
\For {$i=1$ to $n_p$}
\State ${\bf D}_i$ $\gets$ lower right $i\times i$ submatrix of ${\bf D}$ 
\State $\bm{\Phi}\gets{\bf D}_i$
\State ${\bf y}\gets{\bf d}_i$
\State ${\bf r}_{[1]}\gets{\bf y}$
\State $\bm{\Phi}'\gets\bm{\emptyset}$
\While{$j\leq k$}
\State $i_{[j]}\gets\argmax_j(
|\langle \bm{\phi}_j,{\bf r}_{[j]}\rangle|/\lVert\bm{\phi}_j\rVert^2_2
)$
\State $\bm{\Phi}'\gets[\bm{\Phi}',\bm{\phi}_{i_{[j]}}]$
\State ${\bf x}_{[j]}\gets\argmin_{\bf x}:\lVert\bm{\Phi}'{\bf x}-{\bf y}\rVert_2^2$
\State ${\bf r}_{[j]}\gets{\bf y}-\bm{\Phi}{\bf x}_{[j]}$
\EndWhile
\State ${\bf x}_i\gets{\bf x}_{[k]}$
\State Compute roots $z_r$ of $A(\omega)$ using ${\bf x}_i$
\State $\lambda_r\gets-\ln(z_r)/T_s$
\State $f_r\gets|\lambda_r|/2\pi$
\State $\zeta_r\gets-{\rm Re}(\lambda_r)/|\lambda_r|$
\State Draw $f_r$'s on the stability diagram for the polynomial order of $i$
\EndFor
\State Extract stable $f_r$ and $\zeta_r$ from the stabilization diagram
\State Compute mode shapes using $f_r$ and $\zeta_r$
\end{algorithmic}
}
\end{algorithm}

\section{Numerical validation} \label{sec:numerical}
To examine the validity of the proposed method, results of numerical experiments are discussed, where the proposed method was applied to numerically-obtained data from harmonic response analysis of a plate using finite element analysis (FEA). 
The method has been tested for two cases: (1) there is no measurement noise in the measured FRFs, and (2) artificial measurement noises are injected to the measured FRFs. 
\subsection{Conditions for the FRF acquisition} 
\begin{figure}[tb]
\centering
\subfigure[Numerical Model]{\includegraphics[width=8.5cm]{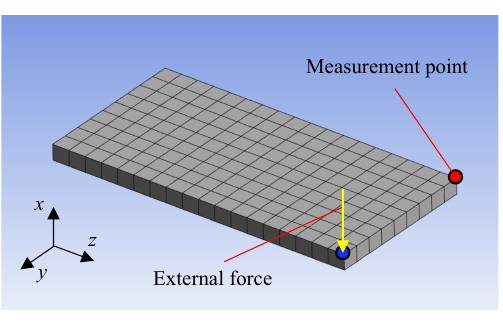}}
\subfigure[FRF evaluated at the measured point]{\includegraphics[scale=.9]{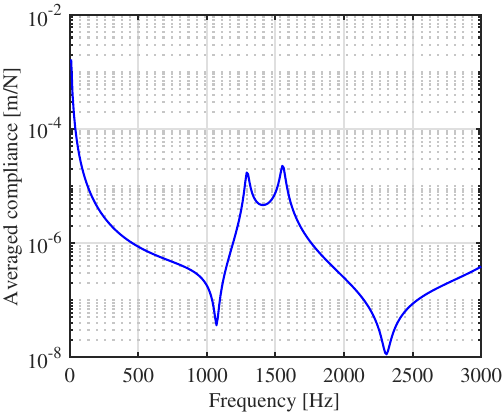}}
      \caption{Numerical model of a plate and its frequency response function}
      \label{fig:study0_model}
\end{figure}

The plate model targeted for the analysis along with the excitation and measurement points is shown in \figref{fig:study0_model} where the dimensions of the model are $200{\rm mm} \times 100 {\rm mm} \times 10 {\rm mm}$, and its material model is structural steel where the density is 7850~kg/m$^3$, Young's modulus is 200~GPa, and Poisson's ratio is 0.3. Harmonic excitation with a unit amplitude was applied in the negative $x$-direction at one of the corner points. 
The structural damping of 0.02 was applied. Note that this would yield the modal damping ratio of 0.01 at the resonant frequencies. 
The FRF of the compliance at the measurement point was obtained from the computed displacement. The frequency range of the harmonic response was set from 10 to 3000 Hz. The boundary conditions were free support. With these conditions, the FEA was conducted using ANSYS 2023R1. 

\begin{figure}[tb]
\centering
	\subfigure[Mode 1 (1292 Hz)]{\includegraphics[width=7cm]{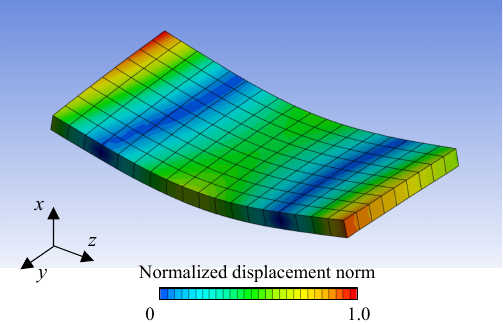}}
	\subfigure[Mode 2 (1553 Hz)]{\includegraphics[width=7cm]{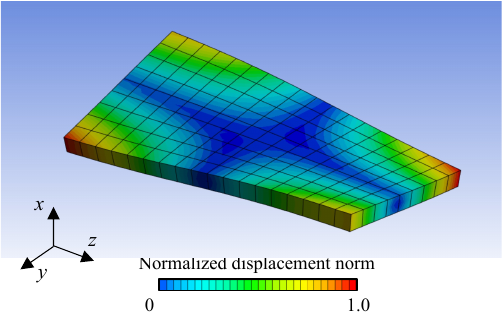}}
      \caption{Vibration modes of the model corresponding to the two resonant peaks. Results are obtained by FEM.}
      \label{fig:study0_modeshape}
      \label{fig:study0_FRF}
\end{figure}

The FRF of the measurement point is presented in \figref{fig:study0_model}(b). 
As can be seen, there are two distinct peaks at 1292 Hz and 1553 Hz. The corresponding response shapes at the resonance peaks are shown in \figref{fig:study0_modeshape}. The Mode 1 shown in \figref{fig:study0_modeshape}(a) is the first out-of-plane bending vibration mode whereas the Mode 2 shown in \figref{fig:study0_modeshape}(b) is the first torsional mode, respectively. 
In the following analysis, the proposed method is applied to the obtained FRF and consider the case where there are no noise in the data, and the case where there are some degree of noise in the data.
\subsection{Result without measurement noise} 
\begin{figure}[h]
    \centering
    \subfigure[Conventional LSCF method]{
        \includegraphics[width=0.45\columnwidth]{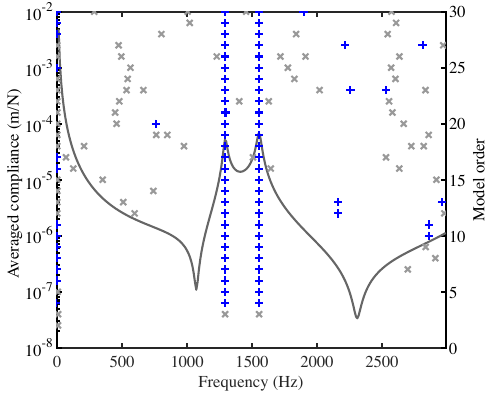}
        \label{fig:study0_conventional_method}
    }
    \subfigure[Proposed method]{
        \includegraphics[width=0.45\columnwidth]{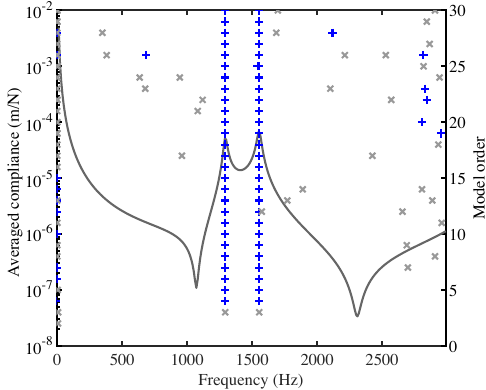}
        \label{fig:study0_proposed_method}
    }
    \caption{Comparison of stability diagram obtained from the conventional LSCF method and the proposed method. ${\color{blue}{\bm{+}}}$: stable consistent pole, $\times$: stable spurious pole.}
    \label{fig:study0_stabilitydiagram}
\end{figure}

\begin{figure}[!tbp]
    \centering
    \includegraphics[width=0.5\linewidth]{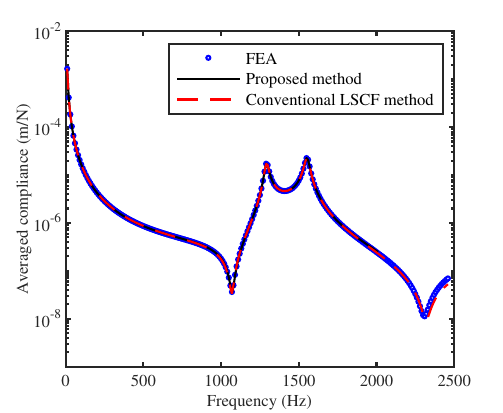}
    \caption{Result of Curve fit when both methods are applied to the FRF obtained from the numerical experiment}
    \label{fig:study0_curvefit}
\end{figure}

\begin{table}[h]
\centering
\caption{Comparison of modal parameters derived by both method from numerical experiment without noise}
\label{tab:study0_modal_parameters}
\begin{tabular}{cccccc}
\toprule
\multirow{2}{*}{\shortstack{Mode \\ number}} & \multicolumn{2}{c}{Natural frequency [Hz]} & \multicolumn{2}{c}{Modal damping ratio}  \\
\cmidrule(lr){2-3} \cmidrule(lr){4-5}
 & Conventional method & Proposed method & Conventional method & Proposed method & \\
\midrule
1 & 1292.4 & 1292.4 & 0.0100 & 0.0099 \\
2 & 1553.8 & 1553.8 & 0.0100 & 0.0100  \\

\bottomrule
\end{tabular}
\end{table}
The proposed method was applied to the data obtained as described above, and the result is compared with that obtained by the conventional method. The calculation results represented as the stability diagram are shown in ~\figref{fig:study0_stabilitydiagram} where \figref{fig:study0_stabilitydiagram}(a) was obtained by the conventional LSCF method whereas \figref{fig:study0_stabilitydiagram}(b) was obtained by the proposed method. For both cases, the maximum polynomial order was set to $n_p=30$. 
It is noted that only the stable poles, or the poles with positive damping ratios of the polynomial with the order indicated by its model order on the right axis are shown in the stability diagram, where a ``${\color{blue}\bm{+}}$'' symbol indicates a pole that consistently appears regardless of the model order and a ``${{\times}}$'' symbol indicates a pole that inconsistently appears depending on the model order, or possibly a spurious pole. The inconsistency of the poles between the model orders is determined whether a pole of the polynomial with a certain model order also exists for a lower order polynomial within a pre-set threshold value, which was set to 1\%. In this paper, the former pole is referred to as a stable consistent pole whereas the latter pole is referred to as a stable spurious pole, respectively. 

From \figref{fig:study0_stabilitydiagram}, we can see that the proposed method produces less number of stable spurious poles than the conventional LSCF method. This means that the stable spurious poles have become unstable poles with negative damping ratios, resulting in disappearing in the stability diagram. 
Figure~\ref{fig:study0_curvefit} shows the results of the curve fit of the FRFs based on the derived modal parameters. As can be seen, the proposed method produces the FRF curve as accurate as the conventional LSCF method. 
\tabref{tab:study0_modal_parameters} compares the results of natural frequencies and damping ratio derived by both methods. Note that the ground truth of the modal damping ratio is 0.01 because the structural damping ratio was set to 0.02 in the harmonic response analysis. From the results in this table, it can be seen that the proposed method derives results as accurate as those of the conventional method. 
\subsection{Result with measurement noise} 
\begin{figure}[!tbp]
    \centering
    \subfigure[Conventional LSCF method]{
        \includegraphics[width=0.45\columnwidth]{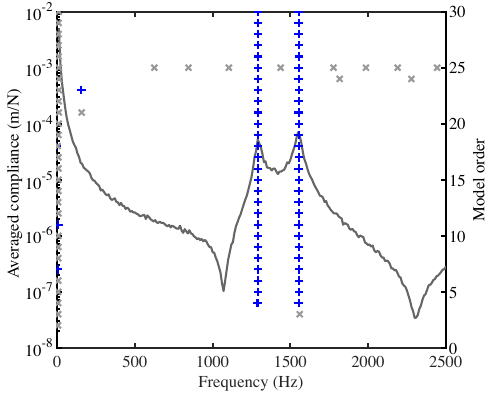}        
        \label{fig:study0_noise_conventional_method}
    }
    \hfill
    \subfigure[Proposed method]{
                \includegraphics[width=0.45\columnwidth]{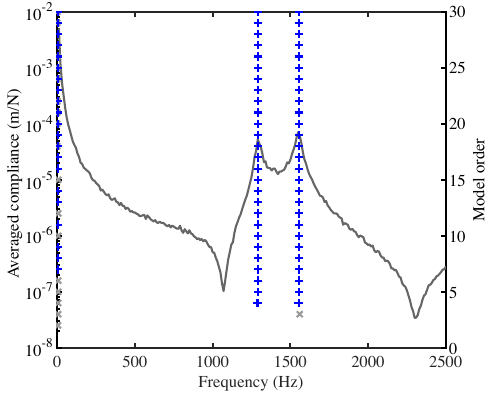}
        \label{fig:study0_noise_proposed_method}
    }
    \caption{Comparison of the stability diagrams when both methods are applied to the FRF obtained from the numerical experiment with noise {\color{black}{($\alpha=0.05$)}}. ${\color{blue}{\bm{+}}}$: stable consistent pole, $\times$: stable spurious pole.}
    \label{fig:study0_noise_stability_diagram}
\end{figure}

\begin{figure}[!tbp]
    \centering
    \includegraphics[width=0.5\linewidth]{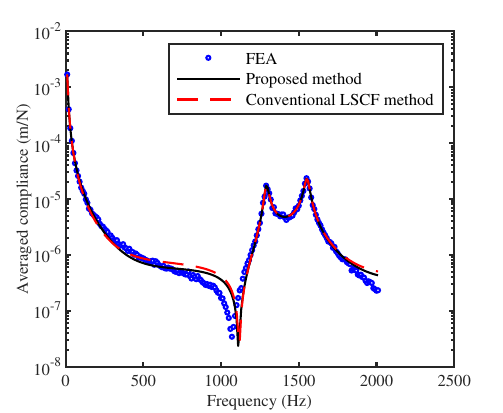}
    \caption{Result of curve fit when both methods are applied to the FRF obtained from the numerical experiment with noise {\color{black}{($\alpha=0.05$)}}}
    \label{fig:study0_noise_curvefit}
\end{figure}
\begin{figure}[!tbp]
    \centering
    \subfigure[Conventional LSCF method]{
        \includegraphics[width=0.45\columnwidth]{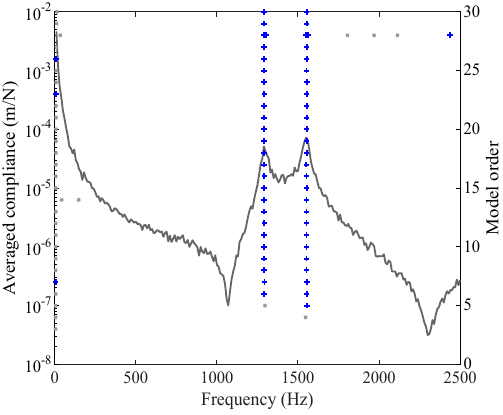}                
        \label{fig:study0_noise_conventional_method_2}
    }
    \hfill
    \subfigure[Proposed method]{
                \includegraphics[width=0.45\columnwidth]{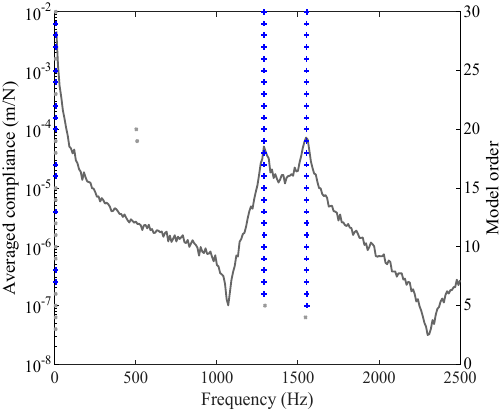}                
        \label{fig:study0_noise_proposed_method_2}
    }
    \caption{Comparison of the stability diagrams when both methods are applied to the FRF obtained from the numerical experiment with noise {\color{black}{($\alpha=0.1$)}}. ${\color{blue}{\bm{+}}}$: stable consistent pole, $\times$: stable spurious pole.}
    \label{fig:study0_noise_stability_diagram_2}
\end{figure}

\begin{figure}[!tbp]
    \centering
    \includegraphics[width=0.5\linewidth]{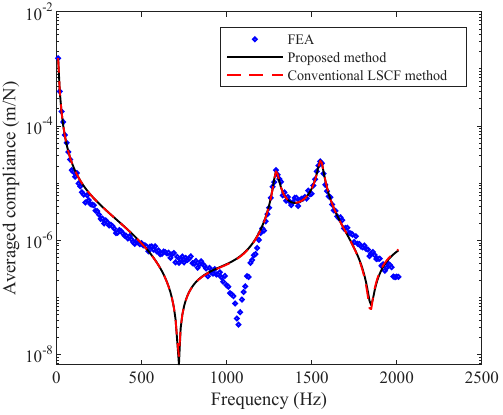}
    \caption{Result of curve fit when both methods are applied to the FRF obtained from the numerical experiment with noise {\color{black}{($\alpha=0.1$)}}}
    \label{fig:study0_noise_curvefit_2}
\end{figure}

\begin{table}[h]
\centering
\caption{Comparison of modal parameters derived by both method from numerical experiment with noise {\color{black}{($\alpha=0.05$)}}}
\label{tab:study0_noise_modal_parameters}
\begin{tabular}{cccccc}
\toprule
\multirow{2}{*}{\shortstack{Mode \\ number}} & \multicolumn{2}{c}{Natural frequency [Hz]} & \multicolumn{2}{c}{Modal damping ratio}  \\
\cmidrule(lr){2-3} \cmidrule(lr){4-5}
 & Conventional method & Proposed method & Conventional method & Proposed method & \\
\midrule
1 & 1291.8 & 1291.7 & 0.0103 & 0.0101 \\
2 & 1553.8 & 1553.8 & 0.0100 & 0.0099  \\ \hline
\end{tabular}
\end{table}

\begin{table}[h]
{\color{black}{
\centering
\caption{Comparison of modal parameters derived by both method from numerical experiment with noise ($\alpha=0.1$)}\label{tab:study0_noise_modal_parameters_2}
\begin{tabular}{cccccc}
\toprule
\multirow{2}{*}{\shortstack{Mode \\ number}} & \multicolumn{2}{c}{Natural frequency [Hz]} & \multicolumn{2}{c}{Modal damping ratio}  \\
\cmidrule(lr){2-3} \cmidrule(lr){4-5}
 & Conventional method & Proposed method & Conventional method & Proposed method & \\
\midrule
1 & 1291.3 & 1291.2 & 0.0106 & 0.0107 \\
2 & 1553.7 & 1553.3 & 0.0096 & 0.0095  \\ \hline
\end{tabular}
}}
\end{table}

Next, the proposed method has been applied to the FRF with the intentional measurement noises. 
Denoting the unaltered FRF as $H(\omega)$, the FRF with the measurement noise is defined as 
\begin{equation}
\hat{H}(\omega)=\left\{1\pm\alpha\sigma(\omega)\right\}H(\omega)
\end{equation}
where $\hat{H}({\omega})$ is the FRF with the injected noise, $\alpha$ is the standard deviation of the noise which was set to 0.05 {\color{black}{and 0.1}}, and $\sigma(\omega)$ contains the random numbers with zero mean and standard deviation of 1.00 with normal distribution. {\color{black}{Note that $\alpha=0.05$ and $0.1$ correspond to signal to noise ratio of 400 and 100, respectively.}}
The maximum polynomial order was set to 30 for both LSCF and the proposed method. The results of both methods {\color{black}{for $\alpha=0.05$}} are shown in \figref{fig:study0_noise_stability_diagram} as stability diagrams, and \figref{fig:study0_noise_curvefit} as curve fit of the FRF. 
It can be seen that in both methods, two vibration modes are accurately derived and a good curve fit has been achieved as before section. Furthermore, conventional method produces stable spurious poles in the stability diagram. On the other hand, the proposed method produces results with much smaller number of stable spurious poles. 
{\color{black}{Furthermore, the results for $\alpha=0.1$ are shown in \figref{fig:study0_noise_stability_diagram_2} as stability diagrams, and \figref{fig:study0_noise_curvefit_2} as curve fit of the FRF. As seen in the figures, even with the larger level of measurement noises, 
both methods accurately find the two modes. Also, the proposed method produces less number of stable spurious poles than those produced by the conventional LSCF method. 
}}

As with the previous case, {\color{black}{Tables}}~\ref{tab:study0_noise_modal_parameters}{\color {black}{ and \ref{tab:study0_noise_modal_parameters_2}} show comparisons} of the natural frequencies and damping ratios {\color{black}{for $\alpha=0.05$ and $\alpha=0.1$, respectively}}. Although there are slight deviations compared to the case without noise, {\color{black}{and the level of deviation increases as $\alpha$ increases}}, it can be seen that the vibration modes have been accurately identified. These results indicate that the proposed method works {\color{black}{fairly}} well even with the existence of measurement noises. {\color{black}{However, the identification accuracy of the proposed method decreases as the level of measurement noise increases.}}

In the following three sections, three case studies with real measurement data are presented to demonstrate the effectiveness of the proposed approach.
%

\section{Case study 1: system with low damping level} \label{sec:study1}
First, the applicability of the proposed method to the case where the expected modal damping ratios are low with relatively sharp resonance peaks. 
\subsection{Acquisition of the frequency response functions}
\begin{figure}[h]
    \subfigure[Aluminum rectangular plate with low damping level]{
        \includegraphics[scale=.9]{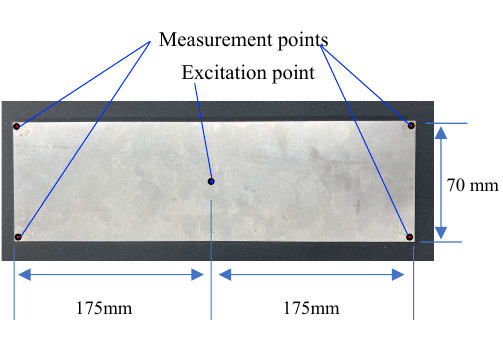}
        \label{fig:study1_model}
    }
    \subfigure[Averaged FRF of the rectangular plate]{
        \includegraphics[scale=.9]{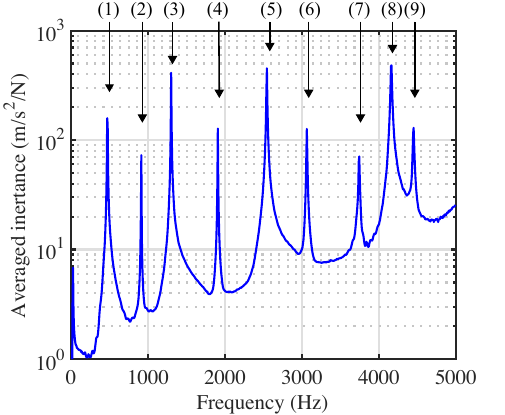}        
        \label{fig:study1_FRF}
    }
    \caption{Experiment model and its frequency response function}
    \label{fig:study1_model_and_FRF}
\end{figure}
The model targeted for the analysis in this section is presented in \figref{fig:study1_model} where the dimension of the rectangular plate is 350mm $\times$ 70mm $\times$ 10mm. The plate is made of an aluminum alloy. A medium-sized impact hammer (086C03, PCB Piezotronics, USA) was used to apply the impulsive forcing at the excitation point indicated in \figref{fig:study1_model}. A tri-axial accelerometers (356A01, PCB, USA) were attached at the four measurement points indicated in \figref{fig:study1_model} to measure the acceleration at those points. 
{\color{black}{The plate was placed on a urethane foam during the test to simulate the free boundary conditions. Note that in the preliminary study, we have conducted experiments with soft suspension of the test specimen using expandable ropes to achieve free-free condition and compared the results with those obtained with the urethane foam support. The errors between the FRFs obtained with the urethane foam support and the ones obtained with the soft suspension were small for the frequency range of interest. Therefore, this support condition was used.}}
An FFT analyzer (OR34J-4, OROS, France) was used for the data acquisition and FRF computations. 
With these conditions, 12 FRFs (three directions per location for four measurement locations) were obtained by the experiments. \figref{fig:study1_FRF} shows the averaged inertance of 12 FRFs obtained from this experiment. 
In the FRF, nine resonance peaks can be seen as indicated by symbols from (1) through (9) in \figref{fig:study1_FRF}. 
\subsection{Results of curve fit} 
\begin{figure}[!tbp]
    \centering
    \subfigure[Conventional LSCF method]{
        \includegraphics[scale=0.9]{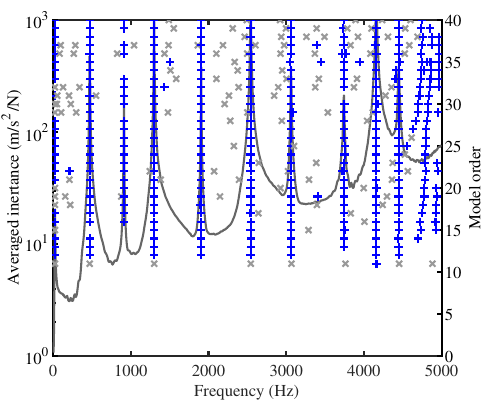}        
        \label{fig:study1_stability_conventional}
    }
    \hfill
    \subfigure[Proposed method]{
        \includegraphics[scale=.9]{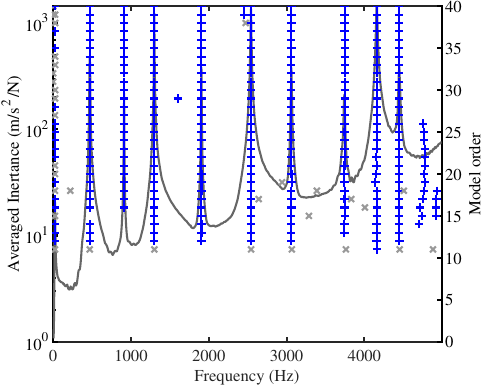}        
        \label{fig:study1_stability_proposed}
    }
    \caption{Comparison of the stability diagrams obtained from the conventional LSCF method and the proposed method. ${\color{blue}{\bm{+}}}$: stable consistent pole, $\times$: stable spurious pole.}
    \label{fig:study1_stability_diagram}
\end{figure}
Both conventional and the proposed methods were applied to the FRF with a fixed polynomial order of 40. The obtained stability diagrams are shown in \figref{fig:study1_stability_diagram}. 
{\color{black}{It is noted that even though the averaged FRF is shown in the figure, the curve fit was conducted for all FRFs.}}
We can see that both methods accurately locate nine vibration modes indicated by the vertically aligned "+" symbols. However, conventional method produced stable spurious poles in the stability diagram. On the other hand, the proposed method produced results with much smaller number of  stable spurious poles. 

\begin{figure}[!tbp]
\centering
\subfigure[LSCF method. Number of stable poles: 554, number of unstable poles: 266.]{\includegraphics[scale=1]{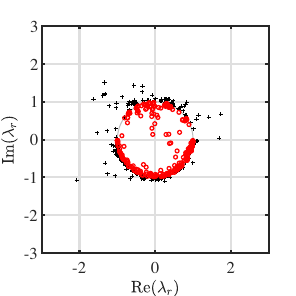}}
\hspace{2em}
\subfigure[Proposed method. Number of stable poles: 369, number of unstable poles: 447.]{\includegraphics[scale=1]{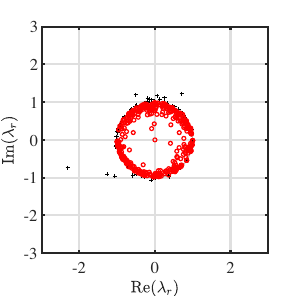}}
\caption{Comparison of the poles in complex plane. ${\color{black}{\circ}}$: inside the unit circle (unstable spurious pole), $+$: outside the unit circle (stable spurious pole)}\label{fig:study1:circle}
\end{figure}
Furthermore, to visualize how the distribution of the poles produced by the methods changes, the obtained poles are plotted in the complex plane and shown in \figref{fig:study1:circle}. Note that all poles produced by the methods for all polynomial orders considered up to the maximum order during the process of drawing the stability diagram are plotted in the figure. 
As can be seen in \figref{fig:study1:circle}(a), there are many poles scattered outside the unit circle produced by the conventional LSCF method. 
Some of these poles correspond to stable spurious poles that appear in the stability diagram, which is not desirable. 
On the other hand, with the proposed method, the number of such stable spurious poles decreases and that of the unstable poles with negative damping ratios increases. 
The numbers of stable and unstable poles produced by both methods are shown in the captions of \figref{fig:study1:circle}. 
Note that some of the degenerate roots produced by the proposed algorithm were eliminated and hence the total number of poles for the proposed method is slightly smaller than that by the LSCF method. 
As can be seen in the figures, the proposed method reduces the number of stable poles while increasing the number of unstable poles. 
This explains why the spurious poles in the stability diagram in \figref{fig:study1_stability_conventional} for the conventional LSCF method were eliminated in \figref{fig:study1_stability_proposed} by the proposed method, because the poles with the negative damping ratios were eliminated and not drawn in \figref{fig:study1_stability_conventional} and \figref{fig:study1_stability_proposed}. 

\begin{figure}[!tbp]
    \centering
    \includegraphics[scale=1]{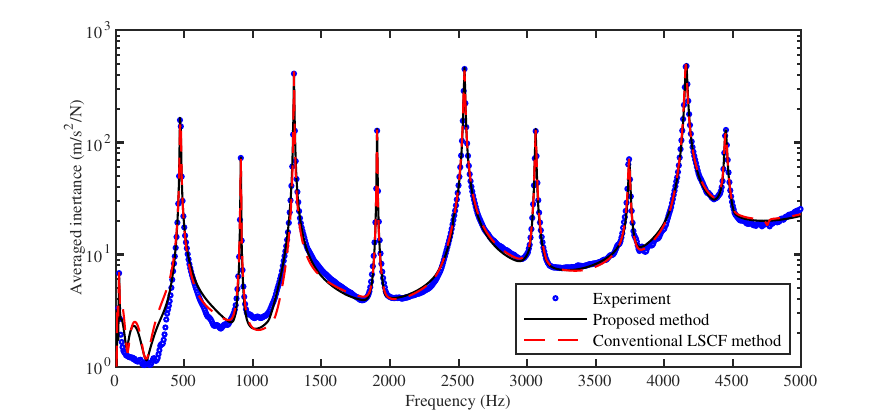}    
    \caption{Result of curve fit when both methods are applied to the FRF obtained from case study 1}
    \label{fig:study1_curvefit}
\end{figure}

Figure~\ref{fig:study1_curvefit} shows the results of curve fit applied to the measured FRF using the conventional method and the proposed method. The proposed method fits the experimentally obtained FRF values as good as the conventional LSCF method. 
The mean squared error (MSE) between the measured FRF and the approximated FRF has been computed for the conventional and the proposed methods. The MSE computed from the results of the conventional method was $16.16 \mathrm {(m/s^2/N)^2}$, while that computed from the proposed method was $4.666 \mathrm {(m/s^2/N)^2}$. This means that the proposed method can derive modal parameters with slightly smaller MSE. 

{\color{black}{In \figref{fig:study1_stability_diagram}(a), the LSCF method identifies poles beyond the mode 9. This is attributed to the following reasons. By inspecting the FRFs over a wider frequency range, there are two strong resonant peaks in the frequency range between 5000 and 6000Hz. Although they are outside the analysis frequency range, their contributions remain in the FRFs below 5000Hz. Since the conventional LSCF method does not include residual terms, to minimize the error near the right edge of the frequency range, the conventional LSCF method needs to express the contributions by the high-order characteristic polynomials. As discussed in Ref.\cite{Verboven2002}, this leads to over-modeling where the accuracy of the least squares fit is improved by the redundant degrees of freedom achieved by the high-order characteristic polynomial. The poles identified by the LSCF method shown in \figref{fig:study1_stability_diagram}(a) can be interpreted as the outcome of such over-modeling of the conventional LSCF method. On the other hand, as can be seen in \figref{fig:study1_stability_diagram}(b), even though the proposed method also identifies some stable consistent poles with relatively low polynomial orders, as the model order increases, the method suppresses the generation of these modes. This is because the proposed method utilizes the characteristic polynomial with sparse set of coefficients, resulting in the reduction of the degrees of freedom of the polynomial to represent the FRF curves. Therefore, the generation of the consistent stable poles are suppressed. Furthermore, as can be seen in \figref{fig:study1_curvefit} the FRFs synthesized by the proposed method is as accurate as the conventional LSCF method. This means that the accuracy of the synthesized FRFs is not deteriorated, yet the generation of stable consistent spurious poles is suppressed. 
}}

\FloatBarrier
\begin{table}[h]
\centering
\caption{Comparison of natural frequencies and damping ratios for the case study 1}
\label{tab:study1_combined}
\begin{tabular}{ccccc@{\hskip 1cm}cccc}
\toprule
\multicolumn{5}{c}{Natural Frequency} & \multicolumn{4}{c}{Damping Ratio} \\
\cmidrule(lr){1-5} \cmidrule(lr){6-9}
\shortstack{Mode \\ number} & LSCF & Proposed & Error [\%] & &
LSCF & Proposed & Error [\%] & \\
\midrule
1 & 472.05 & 471.41 & 0.136 & & 0.00088 & 0.00010 & 12.9 & \\
2 & 911.11 & 911.34 & 0.026 & & 0.00060 & 0.00051 & 15.0 & \\
3 & 1300.2 & 1300.4 & 0.014 & & 0.00112 & 0.00152 & 35.6 & \\
4 & 1905.5 & 1905.9 & 0.023 & & 0.00071 & 0.00082 & 15.1 & \\
5 & 2542.4 & 2543.1 & 0.026 & & 0.00164 & 0.00153 & 6.72 & \\
6 & 3062.3 & 3061.6 & 0.024 & & 0.00137 & 0.00122 & 11.2 & \\
7 & 3743.7 & 3745.8 & 0.058 & & 0.00151 & 0.00139 & 7.51 & \\
8 & 4157.5 & 4159.7 & 0.055 & & 0.00180 & 0.00168 & 6.57 & \\
9 & 4447.4 & 4447.7 & 0.005 & & 0.00196 & 0.00176 & 10.2 & \\
\bottomrule
\end{tabular}
\end{table}
\begin{figure}[h]
    \centering
        \includegraphics[scale=1]{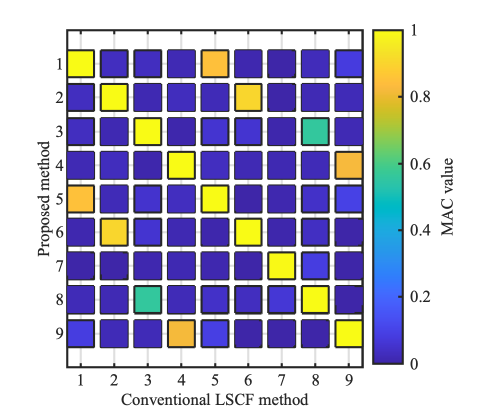}        
      \caption{Comparison of MAC between the mode shapes obtained from the conventional LSCF method and the proposed method for case study 1}  
      \label{fig:study1_MAC}
\end{figure}
\FloatBarrier
 
Next, modal parameters obtained by the proposed methods are discussed. Namely, \tabref{tab:study1_combined} and \figref{fig:study1_MAC} show comparisons of the natural frequencies, damping ratios and Modal Assurance Criterion (MAC)~\cite{Allemang2003} produced by both methods. 
In \tabref{tab:study1_combined}, comparisons of the natural frequencies are shown. We can see that the errors in the natural frequencies are less than 0.2\%. This means that the proposed method produces as accurate natural frequencies as the conventional LSCF method. 
In \tabref{tab:study1_combined}, we can observe that the differences between the damping ratios predicted by the conventional method and those predicted by the proposed method tend to be large. It is known that the damping ratios produced by the LSCF method are error-prone especially when dealing with noisy data~\cite{HoffaitEtAl2019}. 
In \figref{fig:study1_MAC}, the values of MAC between the mode vectors obtained by the conventional LSCF method and those obtained by the proposed method are shown, whose values range from 0 to 1, with values closer to 1 indicating a high similarity between the two modes. From the figure, we can see that the MAC values on the diagonal are above 0.95 or higher. This indicates that the corresponding modes between the conventional method and the proposed method are consistent. Note that some of off-diagonal entries in the MAC have large values. These come from the fact that the mode shapes are measured at a limited number of measurement points and the mode shapes of high-order modes are not well represented by these points. 
From these results, the proposed method accurately identifies the vibration modes for the system with low damping level, and the accuracy is particularly high in deriving the natural frequencies and mode shapes, with some level of errors in the damping ratios. 
\section{Case study 2: system with high damping level} \label{sec:study2}
Next, the proposed method has been applied to a case where high damping level is expected, and the results are discussed. 

\begin{figure}[!tbp]
    \centering
    \subfigure[Rubber rectangular plate with high damping level]{
        \includegraphics[scale=.9]{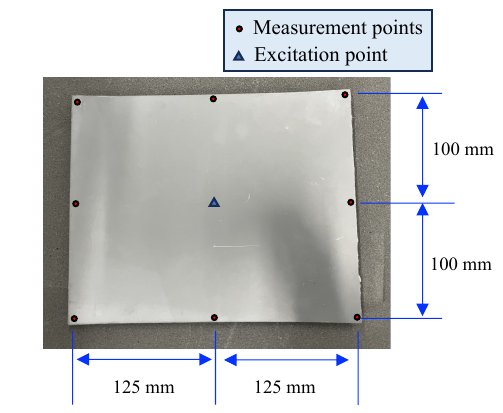}        
        \label{fig:study2_model}
    }
    \hfill
    \subfigure[FRF of the plate evaluated at the measured points]{
        \includegraphics[scale=.9]{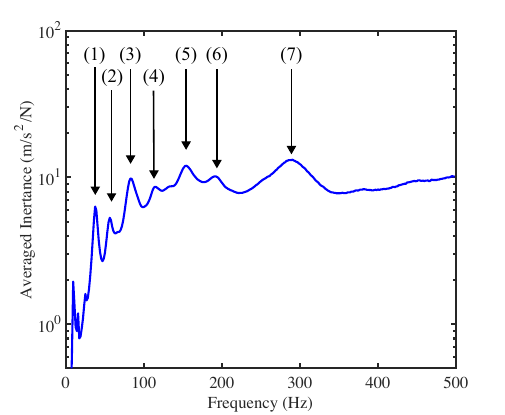}        
        \label{fig:study2_FRF}
    }
    \caption{Experiment model and its frequency response function}
    \label{fig:study2_model_and_FRF}
\end{figure}
\subsection{Acquisition of the FRFs} 
The model targeted for the analysis is presented in \figref{fig:study2_model} where the dimension of the plate is 250mm × 200mm × 10mm. The plate is made of natural rubber. 
When conducting the hammering test, the impact hammer, triaxial accelerometer, and the FFT analyzer used in case study 1 have been used. 
As in the previous case study, the plate was placed on a urethane foam during the test to simulate the free boundary conditions. 
Impulse excitation was applied at the center of the plate along the out-of-plane direction, and the acceleration was measured at the eight points indicated in \figref{fig:study2_model}. 
In total, 24 FRFs were obtained. The frequency range was set from 0 to 500 Hz.  Figure~\ref{fig:study2_FRF} shows the average of 24 FRFs obtained from this experiment. As seen, due to the large damping, the resonant peaks are not as sharp as the ones observed in the previous case study. There are seven distinct resonant peaks indicated in \figref{fig:study2_FRF} and further discussed below. 

\subsection{Results of curve fit} 
\begin{figure}[!tbp]
    \centering
    \subfigure[Conventional LSCF method]{
        \includegraphics[scale=.92]{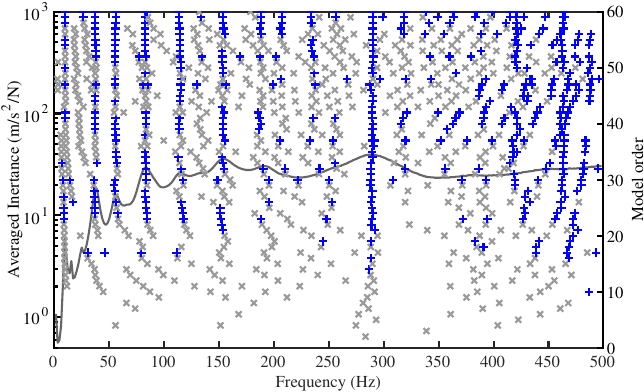}        
        \label{fig:study2_stability_conventional}
    }
    \hfill
    \subfigure[Proposed method]{
        \includegraphics[scale=.92]{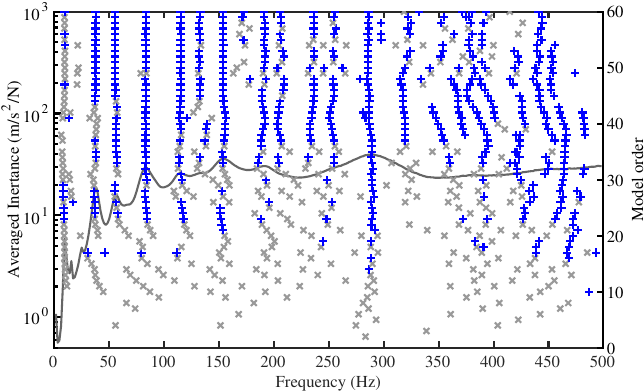}        
        \label{fig:study2_stability_proposed}
    }
    \caption{Comparison of stability diagram when both methods are applied to the FRF obtained from case study 2. ${\color{blue}{\bm{+}}}$: stable consistent pole, $\times$: stable spurious pole.}
    \label{fig:study2_stability_diagram}
\end{figure}
Both the LSCF and the proposed methods have been applied to the obtained FRFs. The maximum polynomial order was set to 60 for both methods. The stability diagrams obtained from both methods are shown in \figref{fig:study2_stability_diagram}. 
As shown in \figref{fig:study2_stability_diagram}, the LSCF method produces many stable spurious poles. Therefore, it is difficult to track the real stable poles of the system. On the other hand, the proposed method produces clearer stability diagram with less number of stable spurious poles. 
{\color{black}{It is noted that the high-order spurious poles are effectively eliminated while the suppression of low-order spurious poles is less significant. This is because the proposed method sparsify the polynomial terms except the largest order term, and hence the effect of the sparsification becomes gradually increases, or the generation of spurious poles tend to be suppressed, as the largest polynomial order increases.}}
\begin{figure}[!tbp]
\centering
\subfigure[LSCF method. Number of stable poles: 1482, number of unstable poles: 348.]{\includegraphics[scale=1]{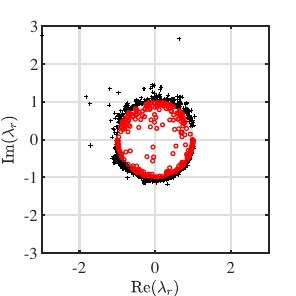}}
\hspace{2em}
\subfigure[Proposed method. Number of stable poles: 1183, number of unstable poles: 647.]{\includegraphics[scale=1]{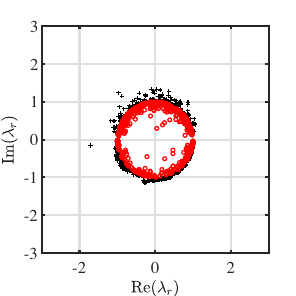}}
\caption{Comparison of the poles in complex plane. ${\color{black}{\circ}}$: inside the unit circle (unstable), $+$: outside the unit circle (stable)}\label{fig:study2:circle}
\end{figure}

Figure.~\ref{fig:study2:circle} shows the distributions of the poles produced by both methods for all polynomial orders considered. As in the previous case study, the stable poles scatter outside the unit circle for the LSCF method, as shown in \figref{fig:study2:circle}(a). On the other hand, the number of unstable poles increases for the proposed method, as shown in \figref{fig:study2:circle}(b) and the number of (visible) unstable poles decreases. The numbers of stable and unstable poles shown in \figref{fig:study2:circle} are shown in the captions of \figref{fig:study2:circle}. Indeed, the proposed method produced less number of stable poles and more unstable poles, which results in reducing the number of spurious unstable poles in the stability diagram in \figref{fig:study2_stability_proposed}. 

\begin{figure}[!tbp]
    \centering
    \includegraphics[scale=.92]{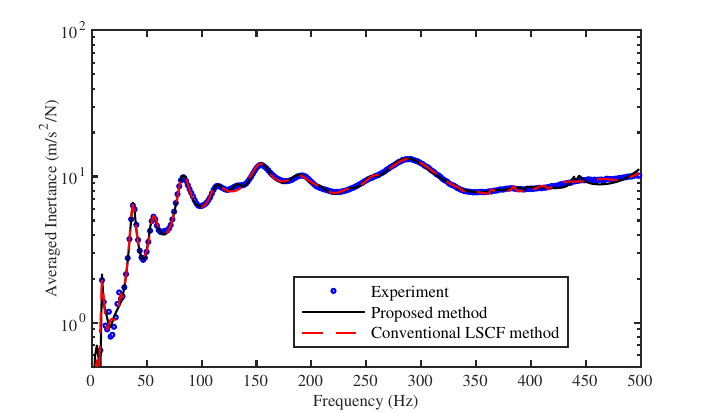}    
    \caption{Result of Curve fit when both methods are applied to the FRF obtained from case study 2}
    \label{fig:study2_curvefit}
\end{figure}

\figref{fig:study2_curvefit} show the results of curve fit using the conventional LSCF and the proposed methods. The MSE values between the measured FRFs and the approximated FRFs computed by the conventional method and the proposed method are $0.0300 \mathrm{(m/s^2/N)^2}$, and $0.0618 \mathrm {(m/s^2/N)^2}$, respectively. This means that the proposed method produced curve fit result as accurate as the conventional LSCF method.

\FloatBarrier
\begin{table}[h]
\centering
\caption{Comparison of natural frequencies and damping ratios for the case study 2}
\label{tab:study2_combined}
\begin{tabular}{ccccc@{\hskip 1cm}cccc}
\toprule
\multicolumn{5}{c}{Natural Frequency [Hz]} & \multicolumn{4}{c}{Damping Ratio} \\
\cmidrule(lr){1-5} \cmidrule(lr){6-9}
\shortstack{Mode \\ number} & LSCF & Proposed & Error [\%] & &
LSCF & Proposed & Error [\%] & \\
\midrule
1 & 37.34 & 37.69 & 0.96 & & 0.0605 & 0.0568 & 6.04 & \\
2 & 56.02 & 55.48 & 0.96 & & 0.0665 & 0.0747 & 12.4 & \\
3 & 83.35 & 83.50 & 0.19 & & 0.0641 & 0.0656 & 2.44 & \\
4 & 113.5 & 115.1 & 1.33 & & 0.0699 & 0.0604 & 13.6 & \\
5 & 154.0 & 154.0 & 0.04 & & 0.0673 & 0.0687 & 2.12 & \\
6 & 190.7 & 190.7 & 1.34 & & 0.0650 & 0.0649 & 0.16 & \\
7 & 289.9 & 286.0 & 1.36 & & 0.0851 & 0.0880 & 3.38 & \\
\bottomrule
\end{tabular}
\end{table}

In \tabref{tab:study2_combined}, natural frequencies obtained by the methods are shown. As seen, the errors in the natural frequencies have increased in comparison with those observed for the case study 1, partially because the resonant peaks are not as sharp as the ones observed in case study 1 because of the damping. However, the errors in the natural frequencies produced by the proposed method are still below 1.5\% and the results are as good as the conventional method. 

Table~\ref{tab:study2_combined} shows the damping ratios produced by both methods. As expected, the damping ratios are larger than the ones observed in the case study 1. Indeed, they are approximately 50 times larger than those for the case study 1. The errors in the damping ratios, however, decreased to an extent in comparison with those observed in the case study 1. This makes sense considering that the algorithm can identify the damping ratios with the dull resonant peaks relatively easier than with sharp resonant peaks, which result in small errors between the conventional and the proposed methods.

\begin{figure}[tb]
    \centering
        \includegraphics[width=0.5\linewidth]{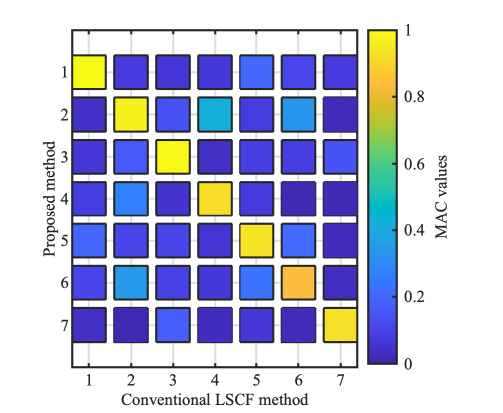}        
      \caption{Comparison of MAC when both methods are applied to the FRF obtained from case study 2}  
      \label{fig:study2_MAC}
\end{figure}

Figure~\ref{fig:study2_MAC} shows the MAC values between the modes produced by the conventional LSCF method and those by the proposed method. As can be seen, the values on the diagonal are 0.9 or higher. This time, the off-diagonal entries are relatively low. This means that the mode shapes produced by the proposed method are consistent with the ones by the LSCF method. 
%

\section{Case study 3: industrial application} \label{sec:study3}
The proposed method has been applied to an industrial application: electric motor stator for a traction drive of electric vehicles. 
The results are discussed in detail. 
\subsection{Acquisition of the FRFs} 
\begin{figure}[!tbp]
    \centering
    \subfigure[Definition of measurement and excitation points]{
        \includegraphics[scale=1]{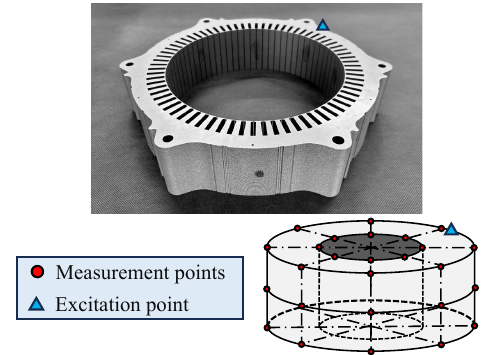}        
        \label{fig:study3_experiment}
    }
    \hfill
    \subfigure[FRF evaluated at the measured points of stator]{
        \includegraphics[width=0.45\columnwidth]{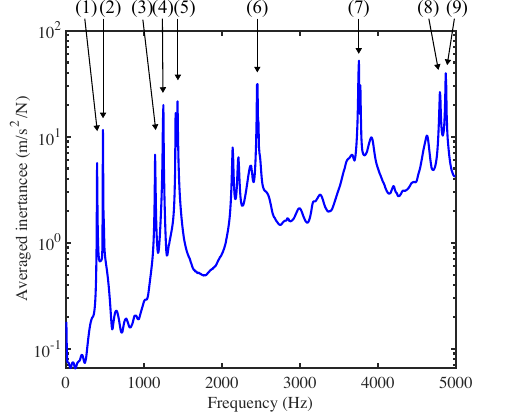}        
        \label{fig:study3_FRF}
    }
    \caption{Experimental model and the averaged measured FRF}
    \label{fig:study3_model}
\end{figure}

The system targeted for the analysis in this study is an electric machine's stator core, whose outer diameter is 268 mm, inner diameter is 198 mm, thickness is 62 mm. The photograph of the stator core is shown in \figref{fig:study3_experiment}. The stator core has been built by stacking hundreds of thin electrical steel sheets that are bonded together to form a single core. 
Detailed specifications of the stator can also be found in Ref.~\cite{SaitoEtAl2025}. From the standpoint of structural dynamics, capturing the modal characteristics of this structure is important yet challenging due to the ambiguities in the boundary conditions on the inter-layer regions between the electrical steel sheets~\cite{SaitoEtAl2016}. 

Figure~\ref{fig:study3_experiment} also shows the measurement and excitation points on the stator core. The excitation point was fixed at a single point on the top perimeter of the stator core and the excitation was applied toward the centripetal direction. 
As shown in the schematic diagram, the triaxial accelerometers were placed at a total of 48 measurement points. Eight equally-spaced locations along the circumferential directions were defined at the intersections of the top, middle, and bottom planes with the inner and outer surfaces of the stator core. The acceleration was measured at radial, circumferential, and axial directions at each measurement point for the impact load at the excitation point, resulting in a total of 144 FRFs. All equipment used for the FRF acquisition including impact hammer, triaxial accelerometers, and the FFT analyzer was the same as the ones used for the previous two case studies. 
The frequency range was set from 0 to 5000 Hz where the modes of interest exist. The stator was placed on a urethane foam as in the previous examples. 
Figure~\ref{fig:study3_FRF} shows the average value of the 144 FRFs obtained from the experiment. 
As can be seen in the FRF, there are many resonant peaks in the frequency range considered. 
To discuss the accuracy of the modal parameters produced by the proposed method, represented dominant peaks observed in \figref{fig:study3_FRF} are selected and labeled as (1) through (9). 
\subsection{Results of curve fit} 
\begin{figure}[!tbp]
    \centering
    \subfigure[Conventional LSCF method]{
        \includegraphics[scale=.98]{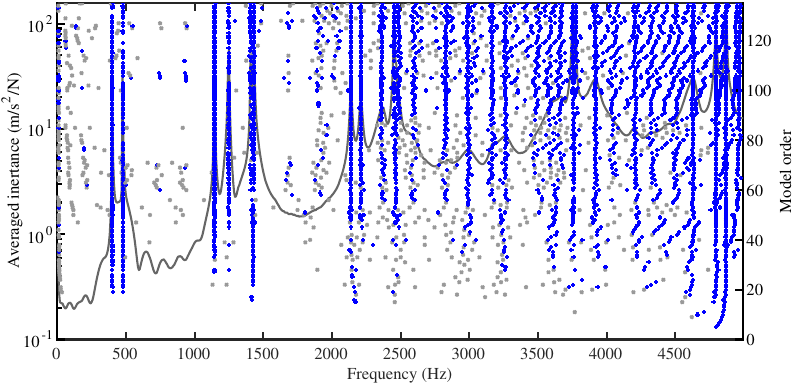}        
        \label{fig:study3_stability_conventional}
    }
    \subfigure[Proposed method]{
       \includegraphics[scale=.98]{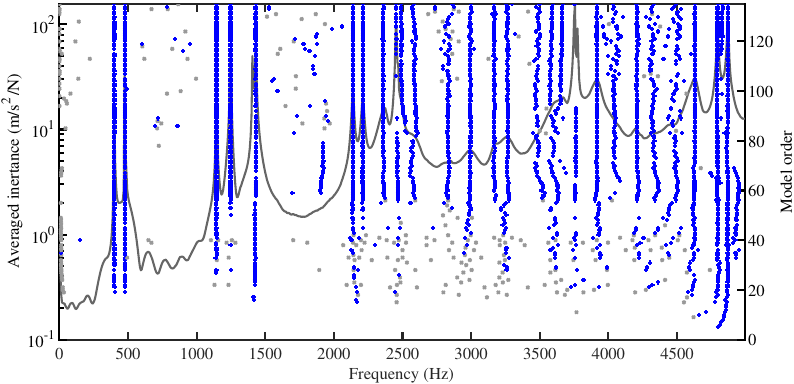}       
        \label{fig:study3_stability_proposed}
    }
    \caption{Comparison of stability diagram when both methods are applied to the FRF obtained from case study 3. ${\color{blue}{\bm{+}}}$: stable consistent pole, $\times$: stable spurious pole.}
    \label{fig:study3_stability_diagram}
\end{figure}
Both the conventional LSCF and the proposed methods have been applied to the measured FRFs. The polynomial order was fixed to 135, which has been increased from the previous case studies because the system is expected to have a larger number of modes in the frequency range considered. 
The obtained stability diagrams are shown in \figref{fig:study3_stability_diagram}. 
As can be seen, the conventional LSCF method produced the stable spurious poles 
that do not appear to correspond to physical modes especially in high frequency range. With this stability diagram, it is hard to distinguish between the stable poles that indeed correspond to vibration modes of the system and the spurious poles that are purely mathematical and do not correspond to any physical modes of the system. 
On the other hand, the proposed method produced results with much smaller numbers of stable spurious poles and stable consistent poles. In particular, the number of stable consistent poles that do not appear to correspond to resonant peaks of the FRF has dramatically decreased. 
Therefore, with the proposed method, we are able to obtain a clearer stability diagram with much smaller numbers of spurious poles, particularly in the high-frequency range, as it was the case for the previous two case studies. 
\begin{figure}[h]
\centering
\subfigure[LSCF method. Number of stable poles: 6146, number of unstable poles: 3034.]{\includegraphics[scale=1]{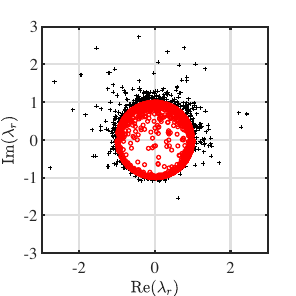}}
\hspace{2em}
\subfigure[Proposed method. Number of stable poles: 3406, number of unstable poles: 5748.]{\includegraphics[scale=1]{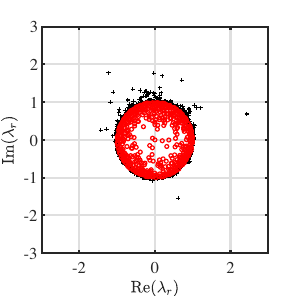}}
\caption{Comparison of the poles in complex plane. ${\color{black}{\circ}}$: inside the unit circle (unstable), $+$: outside the unit circle (stable)}\label{fig:study3:circle}
\end{figure}

Figure~\ref{fig:study3:circle} shows the distributions of the poles obtained by the two methods in complex plane. As it was seen in the previous two case studies, with the LSCF method, there are more stable poles than the unstable poles. With the proposed method, however, the number of unstable poles nearly doubled, while the number of stable poles was reduced by half. This explains that the number of stable poles that appear in the stability diagram shown in \figref{fig:study3_stability_diagram}(b) is much less than the number of stable poles in \figref{fig:study3_stability_diagram}(a) because many stable poles were destabilized and removed in the stability diagram. 

 \begin{figure}[!tbp]
    \centering
    \includegraphics[scale=1]{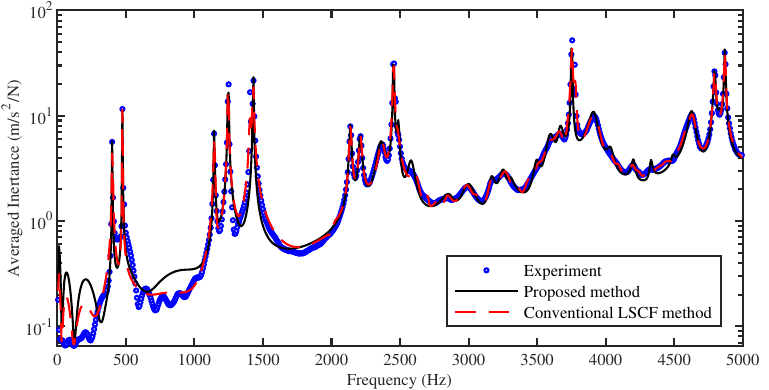}    
    \caption{Result of curve fit when both methods are applied to the FRF obtained from case study 3}
    \label{fig:study3_curvefit}
\end{figure}

Figure~\ref{fig:study3_curvefit} shows the results of curve fit using the conventional and the proposed methods. 
As can be seen, both methods produced accurate results especially at the resonant peaks. 
The MSE between the measured FRF and the approximated FRF of the conventional method shows a value of $0.5213 \mathrm {(m/s^2/N)^2}$, while the proposed method shows a value of $1.7747 \mathrm {(m/s^2/N)^2}$, which is a slight increase from that of the conventional method. 
\FloatBarrier
\begin{table}[h]
\centering
\caption{Comparison of natural frequencies and damping ratios for the case study 3}
\label{tab:study3_combined}
\scriptsize{
\begin{tabular}{cccccc@{\hskip .1cm}cccccc}
\cmidrule(lr){1-6} \cmidrule(lr){7-12}
\multicolumn{6}{c}{Natural Frequency [Hz]} & \multicolumn{6}{c}{Damping Ratio} \\
\cmidrule(lr){1-6} \cmidrule(lr){7-12}
\shortstack{Mode \\ number} & LSCF & Proposed & Error [\%] & {\color{black}{LSCE}}& {\color{black}{Error [\%]}}&& 
LSCF & Proposed & Error [\%] & {\color{black}{LSCE}}&{\color{black}{Error [\%]}}\\
\cmidrule(lr){1-6} \cmidrule(lr){7-12}
1 & 402.30 & 400.89 & 0.352 & {\color{black}{400.81}} & {\color{black}{-0.370}} & &0.00296 & 0.00065 & 78.0 & {\color{black}{0.00252}}&{\color{black}{-14.9}} \\
2 & 475.19 & 475.69 & 0.113 & {\color{black}{475.40}} & {\color{black}{0.044}} & &0.00128 & 0.00020 & 84.3 & {\color{black}{0.00099}}&{\color{black}{-22.7}} \\
3 & 1145.4 & 1144.9 & 0.041 & {\color{black}{1144.7}} & {\color{black}{-0.061}} & &0.00351 & 0.00171 & 51.4 & {\color{black}{0.00217}}&{\color{black}{-38.2}} \\
4 & 1246.2 & 1246.9 & 0.058 & {\color{black}{1247.4}} & {\color{black}{0.092}} & &0.00149 & 0.00045 & 69.9 & {\color{black}{0.00072}}&{\color{black}{-51.7}} \\
5 & 1429.2 & 1429.7 & 0.035 & {\color{black}{1430.4}} & {\color{black}{0.085}} & &0.00283 & 0.00105 & 62.8 & {\color{black}{0.00250}}&{\color{black}{-11.7}} \\
6 & 2453.4 & 2453.0 & 0.015 & {\color{black}{2454.6}} & {\color{black}{0.051}} & &0.00146 & 0.00041 & 72.3 & {\color{black}{0.00132}}&{\color{black}{-9.59}} \\
7 & 3754.1 & 3753.0 & 0.039 & {\color{black}{3758.5}} & {\color{black}{0.116}} & &0.00132 & 0.00022 & 83.2 & {\color{black}{0.00141}}&{\color{black}{6.82}} \\
8 & 4796.6 & 4795.6 & 0.022 & {\color{black}{4796.1}} & {\color{black}{-0.010}} & &0.00159 & 0.00159 & 0.19 & {\color{black}{0.00248}}&{\color{black}{55.9}} \\
9 & 4870.8 & 4869.4 & 0.027 & {\color{black}{4866.9}} & {\color{black}{-0.081}} & &0.00088 & 0.00082 & 6.81 & {\color{black}{0.00197}}&{\color{black}{123}} \\
\cmidrule(lr){1-6} \cmidrule(lr){7-12}
\end{tabular}}
\end{table}

Next, the validity of the obtained modal parameters is discussed. 
The comparisons between the natural frequencies are shown in \tabref{tab:study3_combined}. 
As can be seen, the errors in the natural frequencies are less than 0.5\% for all nine modes considered. This high accuracy in natural frequencies was also observed for the previous two case studies. 
{\color{black}{To further examine the validity of the natural frequencies obtained from the proposed method, the results obtained from the well-established Least Squares Complex Exponential (LSCE) method~\cite{AllemangBrown1998} have also been shown in \tabref{tab:study3_combined}. The errors for the results of the LSCE method are compared with respect to the ones computed from the LSCF method. As seen, the errors are all smaller than 0.5\%. This means that the proposed method produces natural frequencies as accurate as the LSCE method.}}
The damping ratios are shown also in \tabref{tab:study3_combined}. As seen, the damping ratios are relatively small and the results of the proposed method produced some differences from the conventional {\color{black}{LSCF}} method. 
{\color{black}{The damping ratios computed from the LSCE method are also shown in \tabref{tab:study3_combined}. The errors for the LSCE method are computed also with respect to the LSCF method. 
From these results, we can see that the proposed method tends to produce smaller damping coefficients than the conventional LSCF and LSCE methods. 
However, considering that the damping ratios computed by the LSCE method also contain errors with respect to the ones produced by the LSCF method, all these methods are error-prone even if the corresponding natural frequencies are accurately predicted.}}
Even though the values of the damping ratios are as low as the ones for the case study 1, the errors produced by the proposed method are larger than those observed for the case study 1. 
%
{\color{black}{The reason for this is attributed to the perturbations in the coefficients of the characteristic polynomial induced by the measurement noises, and the fact that many of the modes are closely-spaced. This will be further discussed in \secref{discussion} in detail.}}
%

\begin{figure}[!tbp]
    \centering
    \subfigure[Conventional LSCF method]{        \includegraphics[width=0.47\columnwidth]{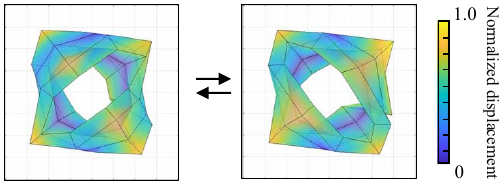}    }
    \subfigure[Proposed method]{        \includegraphics[width=0.47\columnwidth]{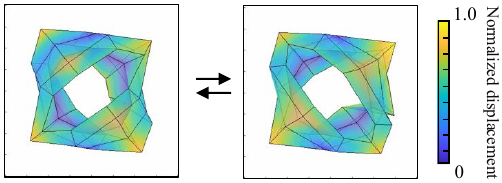}    }

    \caption{Mode shapes for the mode 1. MAC=0.998.}
    \label{fig:modeshape1}
        \vspace{-1em}
\end{figure}
\begin{figure}[!tbp]
\centering
    \subfigure[Conventional LSCF method]{        \includegraphics[width=0.47\columnwidth]{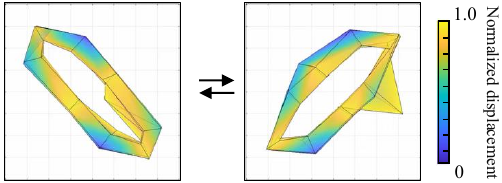}    }
    \subfigure[Proposed method]{        \includegraphics[width=0.47\columnwidth]{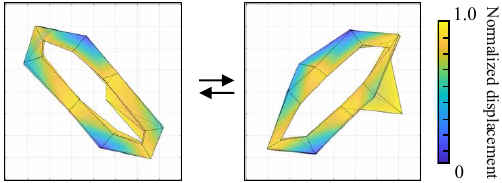}    }

    \caption{Mode shapes for the mode 2. MAC=0.999.}
    \label{fig:modeshape2}
        \vspace{-1em}
\end{figure}
\begin{figure}[!tbp]
    \centering
    \subfigure[Conventional LSCF method]{        \includegraphics[width=0.47\columnwidth]{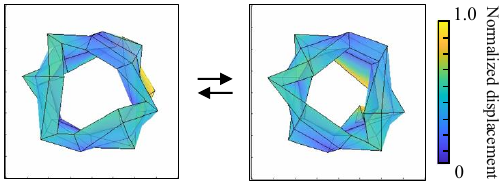}    }
    \subfigure[Proposed method]{        \includegraphics[width=0.47\columnwidth]{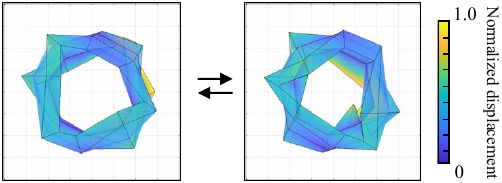}    }    
    \caption{Mode shapes for the mode 3. MAC=0.986.}
    \label{fig:modeshape3}
        \vspace{-1em}
\end{figure}
\begin{figure}[!tbp]
\centering
    \subfigure[Conventional LSCF method]{        \includegraphics[width=0.47\columnwidth]{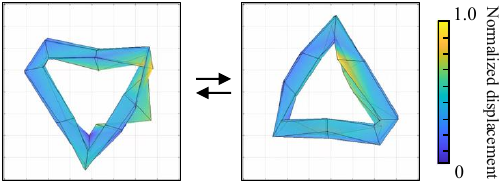}    }
    \subfigure[Proposed method]{        \includegraphics[width=0.47\columnwidth]{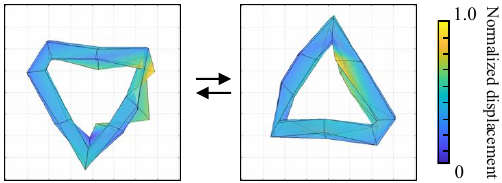}    }
    \caption{Mode shapes for the mode 4. MAC=0.999.}
    \label{fig:modeshape4}
        \vspace{-1em}
\end{figure}
\begin{figure}[!tbp]
    \centering
    \subfigure[Conventional LSCF method]{        \includegraphics[width=0.47\columnwidth]{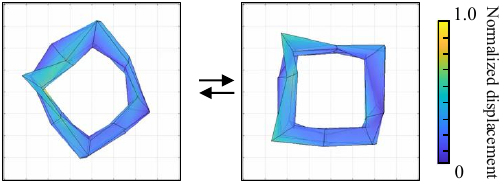}    }
    \subfigure[Proposed method]{        \includegraphics[width=0.47\columnwidth]{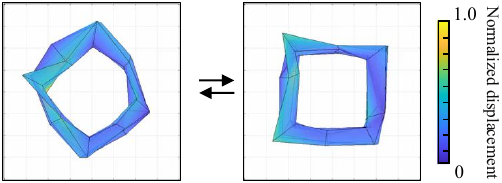}    }
    \caption{Mode shapes for the mode 6. MAC=0.990.}
    \label{fig:modeshape5}    
    \vspace{-1em}
\end{figure}
\begin{figure}[!tbp]
\centering
    \subfigure[Conventional LSCF method]{        \includegraphics[width=0.47\columnwidth]{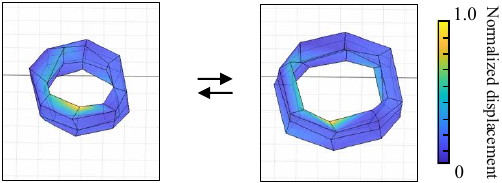}    }
    \subfigure[Proposed method]{        \includegraphics[width=0.47\columnwidth]{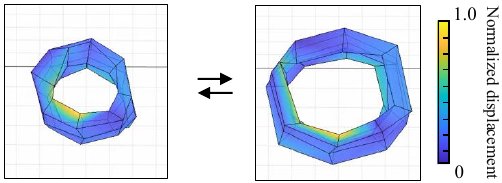}    }    
    \caption{Mode shapes for the mode 8. MAC=0.950.}
    \label{fig:modeshape6}
        \vspace{-1em}
\end{figure}
%

%

\begin{figure}[!tbp]
    \centering
	\subfigure[Proposed method veruss LSCF method]{\includegraphics[width=0.45\linewidth]{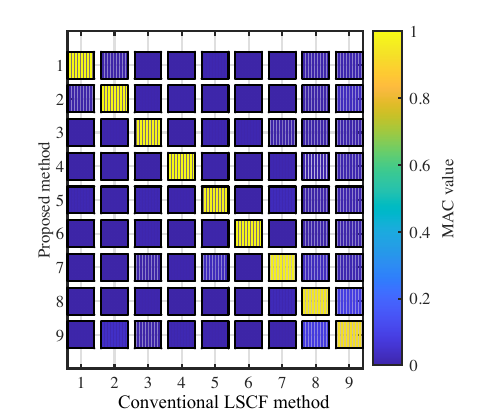}}	
        \subfigure[{\color{black}{Proposed method versus LSCE method}}]{\includegraphics[width=0.4\linewidth]{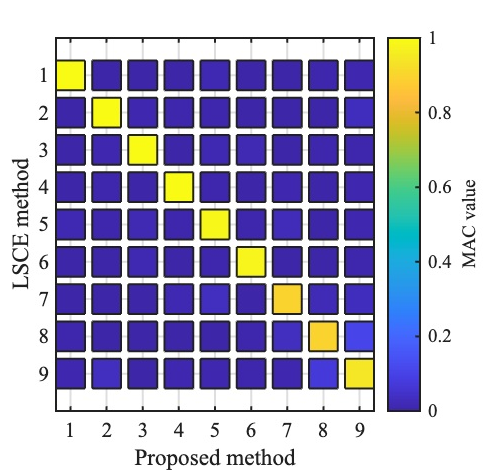}}
      \caption{{\color{black}{Comparison of MAC values for case study 3}}}\label{fig:study3_MAC}
\end{figure}

To visualize and discuss the accuracy of the mode shapes derived by the proposed method, six representative modes were selected from the nine modes previously discussed, and the mode shapes are shown in Figs~\ref{fig:modeshape1} through \ref{fig:modeshape6}. Each figure shows the two moments when the displacement norm becomes the maximum in a vibration cycle. 
{\color{black}{
The mode 1 shown in \figref{fig:modeshape1} shows elliptic deformation with top and bottom layers in axial direction of the core being out-of-phase, where the spatial order in circumferential direction is two whereas the one in axial direction is one. 
The mode 2 shown in \figref{fig:modeshape2} shows elliptic deformation with all layers in axial direction of the core being in-phase, where the spatial order in circumferential direction is two whereas the one in axial direction is zero. 
The mode 3 shown in \figref{fig:modeshape3} shows triangular deformation with top and bottom layers in axial direction of the core being out-of-phase, where the spatial order in circumferential direction is three whereas the one in axial direction is one. 
The mode 4 shown in \figref{fig:modeshape4} shows triangular deformation with all layers in axial directions of the core being in-phase, where the spatial order in circumferential direction is three whereas the one in axial direction is zero. 
The mode 6 shown in \figref{fig:modeshape5} shows square-shaped deformation with all layers in axial directions of the core being in-phase, where the spatial order in circumferential direction is four whereas the one in axial direction is zero. 
The mode 8 shown in \ref{fig:modeshape6} shows deformation of the core in radial direction only where both circumferential and axial spatial orders are zero. Due to the opening and closing nature of the mode shape, it is sometimes referred to as {\it breathing mode-shape 0}, which is known to be the main source of noise and vibration of electric machines especially for the electric vehicles~\cite{HofmannEtAl2014}. 
}}
From these figures, it was found that the mode shapes derived using the proposed method are almost identical to those obtained with the conventional method, demonstrating that the proposed method can derive high-accuracy mode shapes. 
Figure~\ref{fig:study3_MAC}{\color{black}{(a)}} shows the MAC values between the modes derived by {\color{black}{the proposed method and the conventional LSCF method}}. As can be seen in \figref{fig:study3_MAC}{\color{black}{(a)}}, the values of the diagonal terms are close to one, whereas those of the off-diagonal terms are less than 0.1 almost everywhere. 
{\color {black}{Figure~\ref{fig:study3_MAC}(b) shows the MAC values between the modes derived by the LSCE method and the proposed method. As in \figref{fig:study3_MAC}(a), all diagonal terms are close to one and the off-diagonal terms are close to zero, which means that the mode shapes obtained by the proposed method well agree with those obtained by the LSCE method.}}
Based on these results, we can see that the proposed method can also produce accurate mode shapes.\par

{\color{black}{
\section{Discussion}\label{discussion}
Based on the results for the case studies 1, 2, and 3, the damping ratios generated by the proposed method produces errors with respect to the ones computed by the LSCF method. The reason for this is attributed to the variations in the coefficients of the characteristic polynomials. In addition, for the case study 3, the existence of closely-spaced modes also influences the variability in the identified damping ratios, as follows. Denoting the vector of coefficients of the characteristic polynomial as ${\bf a}=[a_{n_p},\dots,a_0]^{\rm T}$ and the $r$th eigenvalue of the characteristic polynomial as $A(z,{\bf {\bf a}})$ as $z_r$, $z_r$ satisfies
\begin{equation}
A(z_r,{\bf a})=0\label{dmp:eq1}
\end{equation}
Now we consider the first order perturbation of $A(z_r,{\bf a})$ with respect to the perturbations in ${\bf a}$ and $z_r$. Namely, denoting the perturbations in ${\bf a}$ as $\delta{\bf a}$, and taking the first perturbation of Eq.~\eqref{dmp:eq1} yields, 
\begin{equation}
\delta A(z_r,{\bf a}) = \sum_{i=0}^{n_p}\frac{\partial A}{\partial a_i}\delta{a}_i + \frac{\partial A}{\partial z_r}\delta z_r = 0
\quad\Rightarrow\quad
\delta z_r=-\left(
\frac{\partial A}{\partial z_r}
\right)^{-1}
\left(
\sum_{i=0}^{n_p}\frac{\partial A}{\partial a_i}\delta{a}_i
\right).
\end{equation}
Considering that the pole in continuous time corresponding to $z_r$ is given as $\lambda_r =-{\rm log}\left(z_r\right)/T_s$, the damping ratio corresponding to $z_r$ is obtained as a function of $z_r$, i.e., 
\begin{equation}
\zeta_r(z_r)=
-\frac{{\rm Re}\left(\lambda_r\right)}{|\lambda_r|}=\frac{\ln|z_r|}{\sqrt{\left(\ln|z_r|\right)^2+\left(\arg(z_r)\right)^2}}
\end{equation}
Therefore, the first-order perturbation of $\zeta_r$ can be obtained as
\begin{equation}
\delta \zeta_r(z_r)=\left(\frac{\partial\zeta_r}{\partial z_r}
\right)\delta z_r=
-\left(\frac{\partial\zeta_r}{\partial z_r}
\right)\left(
\frac{\partial A}{\partial z_r}
\right)^{-1}
\left(
\sum_{i=0}^{n_p}\frac{\partial A}{\partial a_i}\delta{a}_i
\right).
\label{dmp:eq3}
\end{equation}
Using this relationship, the following arguments can be made.

First, equation~\eqref{dmp:eq3} indicates that if $\delta a_i$'s are large, $\delta\zeta_r$ can become large. The perturbations $\delta{a}_i$'s come from two sources. 
One is the perturbation in ${\bf a}$ caused by solving the linear system of equations Eq.~\eqref{eq8} for matrices ${\bf D}_i$'s that contain measurement errors. Another source of the perturbation in $\delta{\bf a}$ is the sparsification of {\bf a} itself, i.e., the proposed method applies large perturbations in the coefficients ${\bf a}$ by making some of them even zeros. Therefore, the proposed method may induce large perturbations in the identified damping ratios. 

Second, considering that 
\begin{equation}
A(z)=\prod^{n_p}_{i=0}(z-z_i), 
\end{equation}
the derivative of $A(z)$ with respect to $z_r$ can be obtained as 
\begin{equation}
\frac{\partial A}{\partial z_r}=-\prod^{n_p}_{i\ne r}(z-z_i)\label{dmp:eq4}
\end{equation}
This indicates that if there are closely-spaced modes near $z_r$, or there exist $i$ such that $z_i\approx z_r$, such as the ones observed for the case study 3, the derivative of $A$ with respect to $z_r$ can be small. It means that the reciprocal of Eq.~\eqref{dmp:eq4} can be large, which leads to the large variations in $\zeta_r$ because of Eq.~\eqref{dmp:eq3}. 

These arguments partially explain why the proposed method produces relatively large errors in the damping ratios, especially for the cases where there are measurement noises. Furthermore, closely-spaced modes may also result in reducing the accuracy of the damping ratios identified by the proposed method. 

To improve the accuracy of the damping ratios of the proposed method, it is important to develop a methodology to quantify the propagation of the variations in the coefficients of the characteristic polynomial to the modal parameters, as in the method proposed in Ref.~\cite{DeTroyer2009}. Such an analysis is beyond the scope of this paper and will be in the future work. 
}}
%
\section{Conclusions} \label{sec:con}
In this paper, we have proposed a novel curve-fit algorithm for the FRFs that incorporates a strategic pole destabilization via OMP into the conventional LSCF method, which results in sparse polynomial coefficients. As a result, the stable spurious poles of the characteristic equations become unstable with negative damping ratios and the resulting stability diagram becomes sparse. 
The method has been applied to numerically-obtained FRFs and to three case studies with physical models including an aluminum plate with small damping level, a rubber plate with high damping level, and a representative industrial application: the motor stator of an electric machine. The results confirmed that non-physical spurious poles produced by the conventional LSCF method were effectively suppressed by the proposed method without significantly compromising the accuracy. 
{\color{black}{The proposed method tends to produce damping ratios with some level of errors in comparison with those predicted by the conventional methods. 
However, the natural frequencies and mode shapes were derived with high accuracy in comparison with those obtained by the conventional LSCF method. 
}}


\section*{Acknowledgement}
The authors would like to thank Honda Motor Co., Ltd. for providing the test piece of the motor stator.


%



\end{document}